\documentclass[twocolumn,showpacs,preprintnumbers,amsmath,amssymb]{revtex4}
\usepackage{epsfig}
\begin{document}
\title{Magnetic Quantum Tunneling in the Single-Molecule Magnet Mn$_{12}$-Acetate}
\author{E. del Barco and A. D. Kent}
\affiliation{Department of Physics, New York University, 4 Washington Place, New
York, NY 10003}
\author{S. Hill}
\affiliation{Department of Physics, University of Florida,
Gainsville, FL 32611-8440}
\author{J. M. North and N. S. Dalal}
\affiliation{Department of Chemistry and National High Magnetic Field
Laboratory, Florida State University,
Tallahassee, FL 32606}
\author{E. M. Rumberger and D. N. Hendrickson}
\affiliation{Department of Chemistry and Biochemistry, University of California
San Diego - La Jolla, CA
92093-0358}
\author{N. Chakov and G. Christou}
\affiliation{Department of Chemistry, University of Florida, Gainsville, FL
32611-7200}
\date{\today}

\begin{abstract}
The symmetry of magnetic quantum tunneling (MQT) in the single
molecule magnet Mn$_{12}$-acetate has been determined by sensitive
low-temperature magnetic measurements in the pure quantum
tunneling regime and high frequency EPR spectroscopy in the
presence of large transverse magnetic fields. The combined data
set definitely establishes the transverse anisotropy terms
responsible for the low temperature quantum dynamics. MQT is due
to a disorder induced locally varying quadratic transverse
anisotropy associated with rhombic distortions in the molecular
environment (2$^{nd}$ order in the spin-operators). This is
superimposed on a 4$^{th}$ order transverse magnetic anisotropy
consistent with the global (average) S$_4$ molecule site symmetry.
These forms of the transverse
anisotropy are incommensurate, leading to a complex interplay
between local and global symmetries, the consequences of which are
analyzed in detail. The resulting model explains: (1) the
observation of a twofold symmetry of MQT as a function of the
angle of the transverse magnetic field when a subset of molecules
in a single crystal are studied; (2) the non-monotonic dependence
of the tunneling probability on the magnitude of the transverse
magnetic field, which is ascribed to an interference (Berry phase)
effect; and (3) the angular dependence of EPR absorption peaks,
including the fine structure in the peaks, among many other
phenomena. This work also establishes the magnitude of the
2$^{nd}$ and 4$^{th}$ order transverse anisotropy terms for
Mn$_{12}$-acetate single crystals and the angle between the hard
magnetic anisotropy axes of these terms. EPR as a function of the angle
of the field with respect to the easy axis (close to the
hard-medium plane) confirms that there are discrete tilts of the
molecular magnetic easy axis from the global (average) easy axis
of a crystal, also associated with solvent disorder. The latter observation
provides a very plausible explanation for the lack of MQT
selection rules, which has been a puzzle for many years.
\end{abstract}

 \maketitle

\section{\label{sec:level1}Introduction}

The origin of magnetic quantum tunneling (MQT) in single-molecule
magnets (SMMs) is important for fundamental reasons as well as
proposed applications, ranging from magnetic data storage
to quantum computing \cite{Tejada,Loss}. A macroscopic
(millimeter-sized) SMM single crystal to a first approximation can
be considered an ensemble of weakly interacting molecules with the
same chemical composition and orientation.  In this regard SMM
single crystals are ideal materials in which to study the quantum
properties of magnetic molecules, as the magnetic response is
amplified by the huge number of molecules forming the crystal. The
environment of the molecules in crystals is also well defined and can
be characterized by techniques such as x-ray diffraction.

SMMs have a predominate uniaxial magnetic anisotropy that
determines an easy magnetic axis for the spin. MQT is due to
interactions that break this axial symmetry and lead to
transitions between magnetic states with opposite projections of
spin on the magnetic easy axis. While this is fundamental to their
quantum dynamics, remarkably only recently have the nature of the
transverse interactions that produce MQT been determined in the
first and most studied SMM, Mn$_{12}$-acetate (henceforth
Mn$_{12}$-ac)
\cite{Friedman,Thomas,Hernandez,Sangregorio,Barra,Perenboom,Bokacheva,Mirebeau,
Hill,Chudnovsky,Chudnovsky2,Cornia,Mertes,Hernandez2,delBarco,delBarco2,Hill2}.
Small modulations in the local environment around the magnetic
cores have been found to be important in MQT
\cite{Cornia,delBarco,delBarco2,Hill2}. In particular, it has also
been shown that in Mn$_{12}$-ac there are a variety of types of
disorder that affect the magnetic properties of the molecules,
including a distribution of solvent microenvironments, $g-$ and
$D-$strain \cite{Park,Hill1,Amigo,Maccagnano,Park1,Parks,Mukhin}
and a distribution of tilts of the easy axes of the molecules
\cite{delBarco3,Hill3}.

In this paper we present high sensitivity magnetometry and high
frequency electron paramagnetic resonance (EPR) studies carried
out on single crystals of Mn$_{12}$-ac. These experimental
techniques and an in-depth analysis of the implications of the
combined data set allow us to determine the symmetry and origin of
MQT. We show that variations in the local solvent environments
around the Mn$_{12}$O$_{12}$ magnetic cores are important to MQT
because they lower the symmetry of the core and lead to tilts of
the magnetic easy axis. Disorder in the molecule's solvent
environment can explain many of the features observed
experimentally. This includes non exponential magnetic relaxation,
the absence of tunneling selection rules, an unusual Berry phase
effect, and multiple and broad EPR absorption peaks.
\begin{figure}
\begin{center}\includegraphics[width=8.8cm]{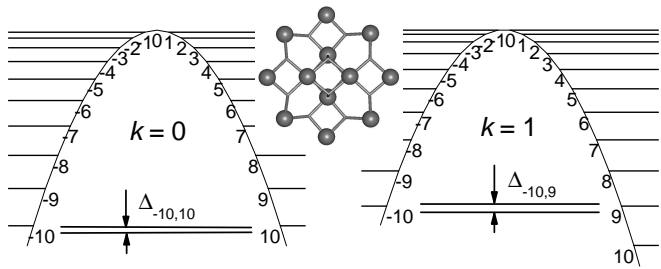}
\vspace{-2 mm}\caption{Energy diagram of the $2S+1=21$ projections
of the spin $S=10$ along the easy magnetic axis of the
Mn$_{12}$-ac molecule for resonances $k=0$ and $k=1$. The tunnel
splitting, $\Delta_{m,m'}$ ($m+m'=k$), is illustrated in both
cases (not to scale). The inset shows the arrangement of Mn ions
looking down the S$_4$ axis of the molecule ($z$-axis). Note that
the spins point up and down along this axis, perpendicular to the
plane of the page.}\vspace{-5 mm}
\end{center}
\end{figure}

The article is organized as follows: In Section II we analyze the
spin Hamiltonian of Mn$_{12}$-ac and discuss the symmetry of MQT
expected based on this Hamiltonian. The effect of a disorder
induced transverse anisotropy combined with intrinsic transverse
anisotropy has interesting consequences for MQT that we explore in
depth. In Section III we describe magnetic
relaxation measurements carried out in the pure quantum regime,
where the MQT is not assisted by thermal activation \cite{Bokacheva}. Section
IV presents high frequency EPR measurements. In Section V we
discuss the implications of our analysis and measurements to the
understanding of MQT.

\section{\label{sec:level1} Tunneling in Mn$_{12}$-ac}

Mn$_{12}$-ac consist of a core of four Mn$^{+4}$ ions
($S=3/2$) surrounded by a ring of eight Mn$^{+3}$ ions (spin
$S=2$) (see fig. 1). These two groups of spins order
ferrimagnetically producing a net spin $S=10$ ($8 \times 2 - 4
\times 3/2 =10$) \cite{Lis,Christou,Regnault}.

The spin Hamiltonian of [Mn$_{12}$O$_{12}$(CH$_{3}$COO)$_{16}$
(H$_{2}$O)$_{4}$]$\cdot$2CH$_{3}$COOH$\cdot$4H$_{2}$O is given by
\begin{equation}
\label{eq.1}{\cal {H}}=-DS_z^2-BS_z^4-g\mu _B\mu_\circ H_zS_z\cos \theta+{\cal
{H_T}}+{\cal {H_A}}+{\cal {H'}}\;,
\end{equation}
The first two terms represent the uniaxial magnetic anisotropy of
the molecule ($D >$ 0 and $B >$ 0). The parameters, $D$ and $B$,
have been determined by high frequency EPR and neutron scattering
experiments \cite{Barra,Hill,Mirebeau}. This spin-Hamiltonian, in
a semiclassical view, describes a double potential well, in which
opposite projections of the spin onto the $z$-axis are separated
by an anisotropy energy barrier $\sim DS^2+BS^4$ ($\sim$ 60 K)
(see fig. 1). The third term is the Zeeman energy due to the
longitudinal component of an external magnetic field,
$H_z=Hcos\theta$, where $\theta$ is the angle between the field
and the easy axis of the molecule. A magnetic field applied along
the $z$-axis shifts the double well potential and the energies of
the projections of the magnetization. At certain values of the
$z$-axis field (resonance fields) the levels $m$ and $m'$ with
antiparallel projections onto the $z$-axis have nearly the same
energy, $H_k\sim kD/g\mu_B = 0.44$ T ($k=m+m'$) (see fig. 1).
\begin{figure}\vspace{+2 mm}
\begin{center}\includegraphics[width=8cm]{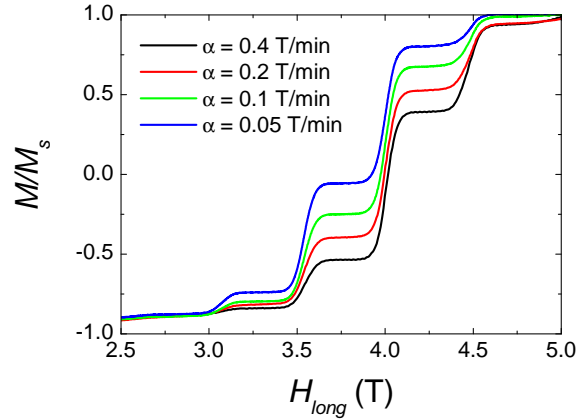}
\vspace{-2 mm} \caption{Magnetization versus field for a single
crystal of Mn$_{12}$-ac for several sweep rates, $\alpha=dH/dt$.
The measurements have been conducted in the pure quantum regime at
$T=$ 0.6 K\cite{Bokacheva}. The observed increases of the
magnetization at regularly spaced field intervals,
$H_k=D/g\mu_B\sim 0.44$ T, correspond to MQT at resonances
$k=6,7,8,9$ and $10$.}\vspace{-5 mm}
\end{center}
\end{figure}

At these resonances, MQT is turned on by interactions that {\it break}
the axial symmetry and mix the levels $m$ and $m'$. These
off-diagonal terms have different origins: (a) $\cal {H_T}$ is the Zeeman
interaction associated with the transverse component of the external
magnetic field; (b) transverse anisotropy terms are included in $\cal {H_A}$;
and (c) $\cal {H'}$ characterizes the transverse components of magnetic fields
due to
inter-molecular dipolar interactions and hyperfine nuclear fields. Off-diagonal
terms in the Hamiltonian lift the degeneracy of the spin-levels
at the resonances creating an energy difference between symmetric
and antisymmetric superpositions of spin-projections that is known
as the tunnel splitting, $\Delta$ (see fig. 1).

Figure 2 shows magnetization curves measured at 0.6 K for
several sweep rates of the longitudinal field, $\alpha=dH/dt$, for
a single crystal of Mn$_{12}$-ac. The abrupt increases of the
magnetization toward the equilibrium magnetization ($M/M_s=1$)
are due to MQT at the resonant fields. It is important to note for our
discussion
that MQT is observed at consecutive resonances. This has important
implications for the understanding of MQT, because transverse
anisotropy terms introduce selection rules and the only interaction
that allows MQT at odd-resonances ($k=1,3,5...$) is a transverse
magnetic field.

\subsection{Transverse Interactions}

The tunneling characteristics depend on the form of the
off-diagonal terms. In this subsection we will examine the consequences of
the off-diagonal terms in the Hamiltonian on MQT. We consider both
transverse magnetic field and anisotropy terms.
\begin{figure}
\begin{center}\includegraphics[width=8.6cm]{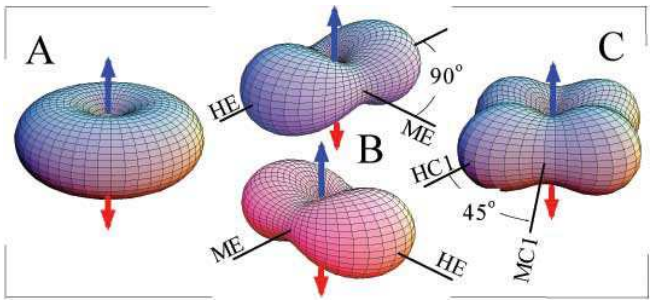}
\vspace{-2 mm}\caption{3D representations of the anisotropy
barrier. (A) Uniaxial anisotropy barrier in the absence of
transverse anisotropy terms. The colored arrows represent the
preferred orientation of the spin along the $z$-axis. (B)
Anisotropy barriers corresponding to opposite signs of second
order anisotropy, $\pm E(S_x^2-S_y^2)$. The lines represent the
hard (H) and medium (M) axes which are separated by 90 degrees. A
change in the sign of $E$ corresponds to a rotation of the
transverse magnetic axes by 90 degrees. (C) Anisotropy barrier due
to a fourth order anisotropy, $C(S_+^4+S_-^4)$. In this case,
there are two hard and two medium transverse axes separated by 45
degrees. A change of the sign of $C$ produces a 45 degree rotation
of the transverse magnetic axes.}\vspace{-8 mm}
\end{center}
\end{figure}

{\it Magnetic Fields.} The simplest expression for the off diagonal part
of the Hamiltonian of Eq. (1), involving a transverse magnetic field,
$H_T$, is
\begin{equation}
\label{eq.2}{\cal {H_T}}=-g\mu_BH_T(S_xcos\phi+S_ysin\phi)\;.
\end{equation}
This represents the Zeeman energy for a field in the $x$-$y$ plane
at an angle $\phi$ with respect to the $x$-axis. The tunnel
splitting, $\Delta$, is very sensitive to this field,
$H_T=\sqrt{H_x^2+H_y^2}$. The dependence of $\Delta$ on $H_T$ for
small transverse magnetic fields ($H_T \ll H_D=2DS/g\mu_B$, the
anisotropy field) is given by \cite{Garanin},
\begin{equation}
\label{eq.3}\Delta_k(H_T)=g_k
\left(\frac{H_T}{H_D}\right)^{\xi_k}\;,
\end{equation}
where
$g_k=\frac{2D}{[(2S-k-1)]^2}\times\sqrt{\frac{(2S-k)!(2S)!}{k!}}$
and $\xi_k=2S-k$. The power law dependence of $\Delta$ on $H_T$
causes the tunnel splitting to vary by many orders of magnitude
for transverse fields in the range of a few Tesla. This allows the
study of MQT with a wide range of experimental
techniques that go from quasi-static magnetization measurements
($\Delta/h \sim$ Hz) to
high frequency EPR experiments ($\Delta/h \sim$ 100 GHz)
simply by varying the magnitude of the applied transverse field.

{\it Transverse Anisotropies.} Now we examine the effect of transverse
anisotropy terms in the
Hamiltonian. Considering only the lowest order terms, $\cal {H_A}$ has the form
\begin{equation}
\label{eq.4}{\cal {H_A}}=E(S_x^2-S_y^2)+C(S_+^4+S_-^4)\;.
\end{equation}
The first term is the second order anisotropy which is allowed for
SMMs with rhombic symmetry (e.g. Fe$_8$ \cite{Wernsdorfer}
and, as will be shown below, is also present in
Mn$_{12}$-ac due to disorder that lowers the S$_4$ symmetry of the
molecules \cite{Cornia,delBarco2,Hill2}). The second term is
fourth order in the spin operators and is the lowest order
anisotropy allowed with tetragonal symmetry. This form of the
transverse anisotropy has been observed in EPR studies of several
SMMs of tetragonal symmetry such as, Mn$_{12}$-ac \cite{Hill2},
Mn$_{12}$-BrAc and Ni$_4$ \cite{EPRNi4}.
\begin{figure}
\begin{center}\includegraphics[width=8cm]{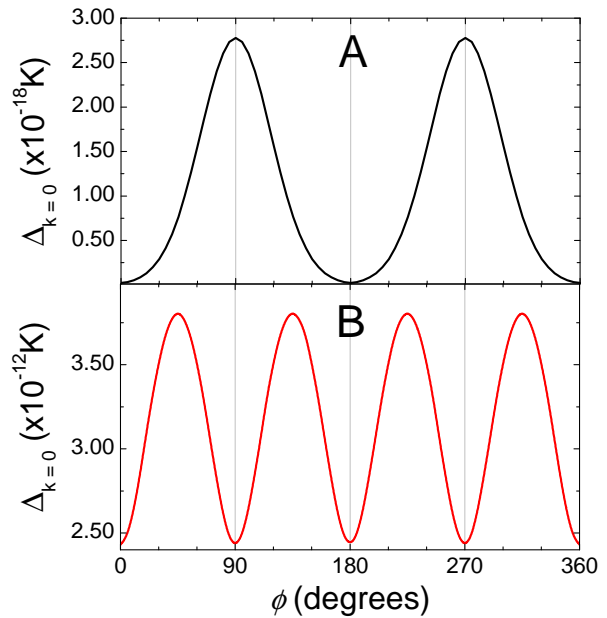}
\vspace{-2 mm}\caption{Tunnel splitting of the ground state
resonance, $k=0$, as a function of the orientation ($\phi$) of a
transverse field of 0.3 T applied in the hard magnetic plane of a
molecule for (A) only second, and (B) only fourth order anisotropy
terms. We have used $D=$ 548 mK, $B=$ 1.17 mK, $E=$ 10 mK and $C=$
0.022 mK in Eq. (1). The graphic clearly shows the 2-fold and
4-fold symmetries imposed by these different anisotropy
terms.}\vspace{-8 mm}
\end{center}
\end{figure}

Let us consider in detail the consequences of each transverse
anisotropy term. In fig. 3 we show the shape of the classical
anisotropy barrier separating antiparallel orientations of the
spin $z$-projections (colored arrows). In the absence of
transverse terms (fig. 3A), the anisotropy barrier is determined
by the uniaxial anisotropy of the molecules (first and second
terms in Hamiltonian of Eq. (1)). These uniaxial anisotropy terms
determine a hard anisotropy plane between opposite orientations of
the spin states along the easy magnetic $z$-axis. Note that this
barrier is isotropic in the x-y plane. In this case, the
tunnel splitting does not depend on the orientation of the transverse
field in the hard plane.

This situation changes in the presence of a transverse anisotropy.
Fig. 3B shows how the anisotropy barrier is modified by a second
order anisotropy term of the form $E(S_x^2-S_y^2)$. This term
introduces one hard and one medium axis in the hard plane (the x-y
plane) that are separated by 90 degrees. That is, for positive $E$
the $x$ is hard and $y$ axis is medium. A change of the sign of $E$ leads to
a 90 degree rotation of these
axes. In this case, the tunnel splitting,
$\Delta$, depends on the azimuthal angle, $\phi$, of the applied transverse
field, $H_T$. A transverse magnetic field applied along the medium
axis produces a larger tunnel splitting than the same field
applied along the hard axis. This leads to an oscillatory
dependence of $\Delta$ on $\phi$ with 2 maxima
and minima separated by 90 degrees (see fig. 4A). A second order
anisotropy also introduces MQT selection rules. In the absence of
a transverse field, this term only allows MQT for resonances that
are even (i.e., $k=2i$ with $i$ an integer).

Fig. 3C shows the modification of the anisotropy barrier in the
presence of a fourth order anisotropy term of the form
$C(S_+^4+S_-^4)$. This anisotropy produces two medium and two hard
axes in the hard plane. The separation between medium and hard
axes is 45 degrees.  Thus the tunnel splitting will have a 4-fold pattern of
oscillations as a function of the angle of a transverse field,
with maxima and minima spaced by a  45 degrees (see fig. 4B). This
term introduces a  selection rule that allows tunneling
transitions from the ground state for resonances that are a
multiple of 4 ($k=4i$). In fig. 4 we show the
expected behavior of the tunnel splitting of the ground state for
resonance $k=0$ versus the angle of the applied transverse field,
$H_T=0.3$ T. The tunnel splitting is calculated using the
Hamiltonian of Eq. (1) with $D=$ 548 mK and $B=$ 1.17 mK and
considering only second order anisotropy ($E=$ 10 mK, fig. 4A) and
only fourth order anisotropy ($C=$ 0.022 mK, fig. 4B).

\subsection{Disorder}

Recent experimental results have shown that MQT in
Mn$_{12}$-ac is modulated by off-diagonal terms that are
generated by disorder. Disorder allows for anisotropy terms that are lower order
in the
spin-operators than those imposed by the average molecule site symmetry in
SMM crystals. Disorder can also lead to tilts of the easy axes
from the crystallographic easy axis
\cite{Mertes,Hernandez2,delBarco,delBarco2,Hill2,delBarco3,Hill3}.
The phenomena that have been explained by disorder can be
summarized as follows:  a) observation of MQT relaxation at
$k$-resonances that are not allowed by the quantum selection rules
imposed by the symmetry of the molecule and b) non-exponential
magnetic relaxation which suggests the existence of a distribution
of tunnel splittings.

Two distinct models of disorder have been proposed. Chudnovsky and
Garanin \cite{Chudnovsky,Chudnovsky2} proposed that random line
dislocations in a crystal lead, via magnetoelastic interactions,
to a lower molecule symmetry and a broad distribution of tunneling
rates. Subsequent magnetic relaxation experiments indeed showed
the existence of a broad distribution of tunneling rates and were
analyzed in terms of this model \cite{Mertes,Hernandez2}. In
contrast, Cornia et al. \cite{Cornia} suggested, based on detailed
x-ray analysis, that variations in the position of the two
hydrogen-bonded acetic acid molecules surrounding the Mn$_{12}$-ac
clusters lead to a discrete set of molecules with lower symmetry
than tetragonal. More recent magnetic relaxation experiments
\cite{delBarco2} and high frequency EPR experiments \cite{Hill2}
have confirmed the latter model, showing a 2-fold symmetry of the
tunnel splitting as a function of the direction of an external
transverse field. Moreover, the observation of a distribution of
transverse fields in Mn$_{12}$-BrAc \cite{delBarco3} and tilts in
Mn$_{12}$-ac \cite{Hill3} suggests that tilts of the easy axes of
the molecules are responsible for the MQT relaxation at odd-$k$
resonances; in the case of Mn$_{12}$-ac, these tilts are caused by
the solvent disorder. We summarize these models and their
consequences below.

{\it Line Dislocations.} Chudnovsky and Garanin
\cite{Chudnovsky,Chudnovsky2} considered a random distribution of
line dislocations with collinear axes to calculate a
representative distribution of second order transverse
anisotropies. They found: a) the corresponding distribution of
tunnel splittings is broad on a {\it logarithmic} scale; b) the
mode of the distribution is at $E=0$; and c), for small
concentrations of dislocations, most molecules are far from the
dislocation cores and the magnetic axes associated with the
transverse anisotropy are oriented randomly. The first point can
explain the non-exponential relaxation found in Landau-Zener
relaxation experiments carried out at several MQT resonances, and
at different longitudinal magnetic field sweep rates
\cite{Mertes}. Furthermore, experiments in which a Mn$_{12}$-ac
single crystal was treated with rapid thermal changes could be
explained in terms of the dislocation model \cite{Hernandez2}.
Point (b) leads to a distribution of tunnel splittings, with a
very long tail towards small values of the tunnel splitting. In a
more recent experimental study, a Landau-Zener method that
involved crossing the same MQT resonance several times permitted
the study of nearly the whole distribution of tunnel splittings in
a Mn$_{12}$-ac single crystal at a single resonance
\cite{delBarco}. The results obtained were compared to the
distribution function proposed by Chudnovsky and Garanin. A
distribution of second order anisotropies with mode at $E=0$ was
not able to explain the relaxation data. More recently, both
magnetic relaxation and high frequency EPR experiments
\cite{delBarco2,Hill2} confirmed the discrete nature of the
distribution of transverse anisotropies and, moreover, the fact
that the magnetic axes of the transverse anisotropy is also
discretely distributed along particular directions in the hard
anisotropy plane of the molecules, contrary to the random
distribution expected from the dislocation model.

{\it Solvent Disorder.} From x-ray diffraction studies Cornia et
al. \cite{Cornia} showed that the fourfold symmetry of
Mn$_{12}$-ac molecules is lowered by disorder of the acetic acid
molecules (2 molecules with 4 possible sites). This gives rise to
a set of six different molecules (4 of them with $E\neq 0$), with
$7/8$ or $88 \%$ of molecules thus having a non-zero second order
transverse anisotropy. The fact that the second order anisotropy
is generated by disorder in the solvent molecules implies that
both the abundance and magnitude of the E for each isomer could
depend on the synthesis process as well as on the solvent losses
that a particular crystal experienced prior to measurement. From
x-ray data Cornia et al. calculated the E parameters for the
different isomers \cite{Cornia}. These should be taken as
estimates, since the data were obtained from one crystal and the
E-parameters were computed using an empirical model. Nonetheless,
this model represents an important step in the understanding of
MQT in Mn$_{12}$-ac.

The relevant points of the solvent disorder model can be
summarized as follows: a) there is a discrete set of $E$ values in
a sample. This means that the distribution of tunnel splittings is
a discrete function with peaks at several $E \neq 0$ values. b)
There are equal populations of isomers having opposite signs of
the second order anisotropy $\pm E$. This means that the medium
and hard axes associated with each $E$ value will be discretely
distributed over the hard plane of the crystal with a separation
of 90 degrees. And c) the hard axes of the second order anisotropy
are rotated away the hard/medium axes corresponding to the fourth
order anisotropy; this rotation angle depends on the precise
details of the solvent structure, and is $\sim$30 degrees for
Mn$_{12}$-ac.. The first point has been confirmed by high
frequency EPR experiments by the observation of discrete peaks in
the absorption spectra that correspond to different discrete $E$
values and have the angular dependence expected from this model
\cite{Hill2}. The second point has been confirmed through magnetic
relaxation measurements in which a portion of molecules having
different signs of $E$ have been studied \cite{delBarco}. The
third point has been observed in EPR experiments and has
implications for the MQT discussed below. We note that both models
suggest the presence of tilts of the easy magnetic axes of the
molecules. These latter results, which have recently been
confirmed via density functional calculations \cite{Pederson},
will be presented in detail in the experimental sections of this
paper.

\subsection{The Effect of Disorder on MQT}

In this section we will present model calculations of the
dependence of the tunnel splitting  on the angle and the magnitude
of an external transverse field taking into account the presence
of both second and fourth order anisotropies. These terms must be
considered on an equal footing because they produce comparable
tunnel splittings. This is seen as follows. In perturbation theory,
the splitting between levels $m$ and $m'$ associated with the
second order anisotropy is approximately  $\Delta_E\sim
D(E/2D)^{(m'-m)/2}$, while that associated with the fourth order
anisotropy is $\Delta_C\sim D(C/2D)^{(m'-m)/4}$. Thus, provided
$E^2 \sim DC$ these are comparable in magnitude, {\it independent} of
$m$ and $m'$. This is  the case for Mn$_{12}$-ac, as
$E\sim10^{-2}$ K, $C\sim10^{-5}$ K and $D\sim1$ K. For this reason
there is an intricate interplay between these anisotropy terms
that also occurs over a broad range of transverse magnetic fields,
and can thus be probed both in EPR and MQT studies.

In Mn$_{12}$-ac these 2nd and 4th order transverse anisotropy terms have a
different
origin. The fourth order transverse anisotropy term is imposed by
the symmetry of the molecule, while the second
order term is generated
by a distribution of solvent micro-environments.
However, there is no reason to assume that these anisotropies are
commensurate since they come from different sources; one is
associated with the ideal S$_4$ symmetry, the other is disorder
induced and lowers this symmetry. Therefore, we will consider a misalignment
angle,
$\beta$, between the magnetic axes associated with each transverse anisotropy
term.

We write the transverse anisotropy part, $\cal{H_A}$, of the
Hamiltonian (Eq. (4)) as follows:
\begin{equation}
\label{eq.5} E(e_1(S_x^2-S_y^2)+e_2(S_x\cdot S_y+S_y\cdot
S_x))+C(S_+^4+S_-^4)\;,
\end{equation}
where,
\begin{equation}
\label{eq.6} e_1=(cos^2\beta-sin^2\beta)\;\;\; e_2=2cos\beta
sin\beta \;.
\end{equation}
This introduces an angle $\beta$ between hard axes of the second and
fourth order anisotropy terms (see fig. 5), the `$E$' and `$C$' terms
respectively.
\begin{figure}
\begin{center}\includegraphics[width=5cm]{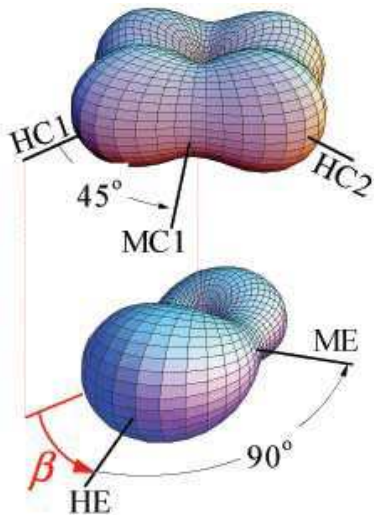}
\vspace{-2 mm}\caption{3-D representation of the anisotropy
barriers corresponding to fourth (upper graph) and second order
(lower graph) anisotropies with an angle $\beta$ between the
corresponding hard axes.}\vspace{-6 mm}
\end{center}
\end{figure}

In the next subsections we will show the effect of having a
misalignment ($\beta$) on the symmetry of MQT and on the
transverse field dependence of the tunnel splitting for different
orientations of the applied field. The calculations have been done
for parameters and conditions close to those shown in the
experimental sections of this article. In the following we
consider in some detail the consequences of the solvent disorder
model as this can adequately explain many of our experimental
observations.

As mentioned, one of the important consequences of the solvent
disorder model is the fact that there must be equal populations of
molecules having opposite signs of $E$. This is because, on average,
Mn$_{12}$-ac crystals have $S_4$ or tetragonal site symmetry. Note that
fig. 5 represents the picture for the molecules having $E>0$. Thus
there will be molecules in the sample with the magnetic axes of
$E$ rotated by 90 degrees with respect to those represented in fig. 5.

{\it Commensurate Transverse Interactions.} We consider the case of
$E>0$ and $\beta=0$ in the presence of large transverse field, as this field
corresponds to
that used in high frequency EPR experiments (section IV). Fig. 6
shows the behavior of the ground state tunnel splitting,
$\Delta_{10,-10}$ (resonance $k=0$), versus the angle, $\phi$, of
an external transverse field of 9 Tesla. The calculations have
been done for $\beta=0$, $D=548$ mK, $B=1.17$ mK, $C=0.022$ mK and
different values of $E$.
\begin{figure}
\begin{center}\includegraphics[width=7cm]{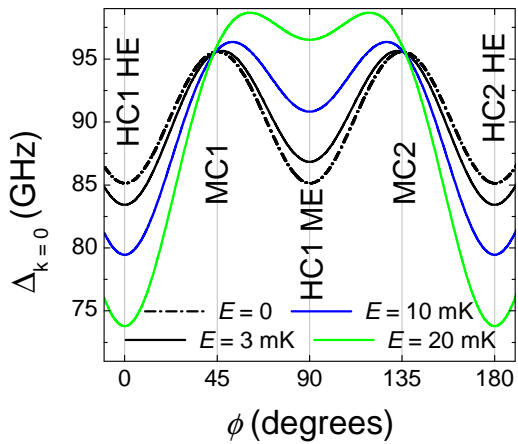}
\vspace{-2 mm}\caption{Behavior of the ground state splitting,
$\Delta_{k=0}$, versus the orientation of a transverse field,
$H_T=9$ T, for $\beta=0$ and different values of $E>0$.}\vspace{-6
mm}
\end{center}
\end{figure}

One interesting result in this figure is the observation of
fourfold pattern  of maxima imposed by the fourth order anisotropy.
There are four maxima in $\Delta$ that occur very close to the
medium axes of $C$ for small values of $E$. Only when $E$ becomes
very large do the maxima move towards the medium axis of $E$ to
give, at high enough $E$-values ($E>>20$ mK), a twofold pattern of
maxima. However, for the parameters estimated with the solvent
disorder model, $E<10-15$ mK, the tunnel splitting is expected to
show fourfold maxima in the direction of the medium axes of $C$.

\begin{figure}
\begin{center}\includegraphics[width=7cm]{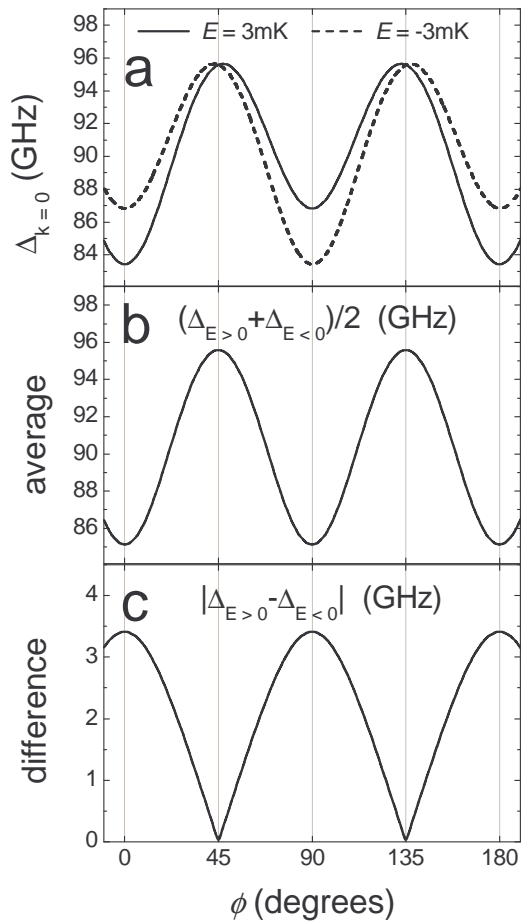}
\vspace{-2 mm}\caption{(a) Ground state tunnel splitting,
$\Delta_{k=0}$, versus the orientation of a transverse field,
$H_T=9$ T, for $\beta=0$ and $E=+3$ mK (solid line) and $E=-3$ mK
(dashed line). (b) Average value of the tunnel splitting assuming
both $E$-signs. (c) Difference between the tunnel splitting values
of $E>0$ and $E<0$.}\vspace{-6 mm}
\end{center}
\end{figure}
Fig. 7a shows the behavior of the ground tunnel splitting,
$\Delta_{k=0}$, for the same conditions as fig. 6, that is for
$\beta=0$, but taking into account the equal population of
molecules having opposite signs of $E$. In this case we have used
$E=E_{av}=\pm3$ mK. The modulation of the fourfold symmetry by $E$
is opposite for different signs of $E$. However, the maxima are
approximately at the same position, independent of the sign of
$E$. In fig. 7b we show the $\phi$ dependence of the average value
of the tunnel splitting $(\Delta_{E>0}+\Delta_{E<0})/2$. This
average exhibits a symmetric four-fold rotation pattern, the same
as would be generated via only a fourth-order anisotropy.
Therefore, a measurement of the average tunnel splitting of a
crystal cannot distinguish between second and fourth order
transverse anisotropies if there are equal populations of
molecules with opposite signs of $E$.

The absolute difference between the tunnel splittings,
corresponding to opposite signs of $E$, is shown in fig. 7c. In
this case, $\beta=0$, the difference is maximal along the hard
axes of $C$ and vanishes along the medium axes of $C$.
Importantly, this provides a means of inferring the presence of a
second-order anisotropy through an appropriate experiment (this
will be shown below and in section IV).

\begin{figure}
\begin{center}\includegraphics[width=7cm]{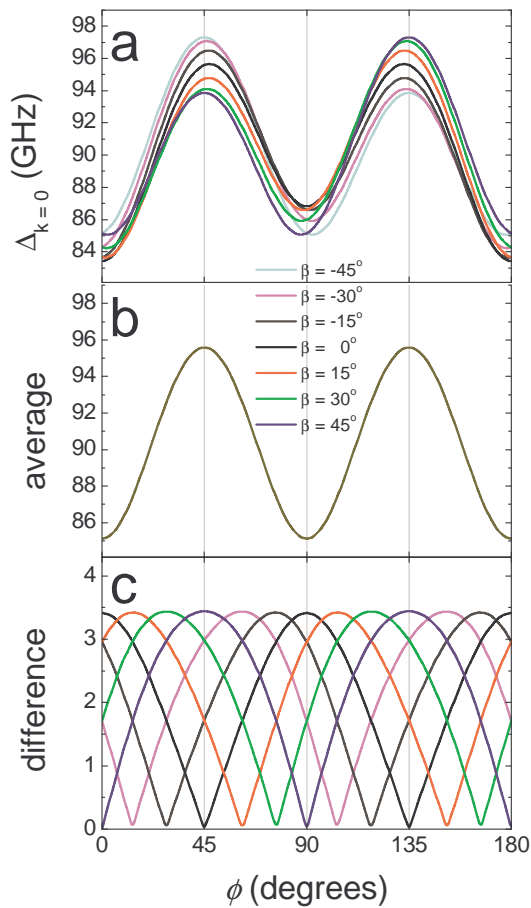}
\vspace{-2 mm}\caption{(a) Ground state tunnel splitting,
$\Delta_{k=0}$, versus the orientation of a transverse field,
$H_T=9$ T, for $E=+3$ mK and different values of $\beta$. (b)
Average value of the tunnel splitting assuming both $E$-signs for
each value of $\beta$; all of the curves lie on top of each other.
(c) Difference between the tunnel splitting values for $E>0$ and
$E<0$ for different $\beta$-values.}\vspace{-6 mm}
\end{center}
\end{figure}
{\it Incommensurate Transverse Interactions.} We now consider the
effect of a misalignment, $\beta\neq0$. Fig. 8a shows the behavior
of the ground tunnel splitting, $\Delta_{k=0}$, versus the angle
of the applied transverse field, $H_T=9$ T for $E=3$ mK and
different values of $\beta$, from $\beta=-45^\circ$ to
$\beta=45^\circ$ in increments of 15 degrees. Note that the black
line, $\beta=0$, is the result presented in fig. 7a.  A
misalignment $\beta\neq0$ generates an asymmetry between the
maxima of the tunnel splitting. For example, for
$\beta=-45^\circ$, the maximum at $\phi=45^\circ$ is bigger than
the maximum at $\phi=135^\circ$. Note that the hard axis of $E$
(HE) for this value of $\beta$ is $\phi=-45^\circ+n180^\circ$
(with $n$ an integer) while the medium axis (ME) is along
$\phi=45^\circ+n180^\circ$. In general, even though the ME axis is
at $\phi=\beta+(2n+1)90^\circ$, {\it the maxima of the tunnel
splitting are in the direction of the medium axes of $C$}. The
second order anisotropy introduces an asymmetric modulation of the
maxima and minima of the tunnel splittings that depends on the
value and the sign of $E$ and on the misalignment angle $\beta$.

Fig. 8b shows that the average value of the tunnel splitting is
independent of the angle of misalignment $\beta$ and has four
maxima. All the results collapse in the same curve. So again a
direct measurement of the average value of $\Delta$ would not give
information about the misalignment between these anisotropies.

Fig. 8c shows the difference between $\Delta_{E>0}$ and
$\Delta_{E<0}$ for the same parameters used in the previous
calculation and different values of $\beta$ from $-45^\circ$ to
$45^\circ$. This difference has four maxima for all the
angles $\beta$. However, the positions
of the maxima are different for different $\beta$ values. In fact,
the position of the maxima depends directly on the value of
$\beta$ as $\phi_{max}=\beta+n90^\circ$.
Consequently, a measurement of this difference would not only
provide the value of $E$ but also give the angle of misalignment
between second and fourth order transverse anisotropies.

We thus note that within the solvent disorder model the
tunnel splitting of resonance $k=0$ (the ground state degeneracy)
at high transverse field values will show a fourfold rotation pattern with
respect the
angle of orientation of the applied field. This symmetry is
modulated by the second order anisotropy. However, equal
populations of molecules with opposite signs of $E$ will give an
average tunnel splitting with four maxima that is
indistinguishable from that corresponding fourth order transverse
anisotropy. The only way to determine the presence of second order
anisotropy is by examining the difference of the tunnel splittings
corresponding to molecules with opposite signs of $E$. The latter
would give the value of $E$ and the misalignment angle between the
second and fourth order anisotropies. We will show how such
measurements are possible via two different techniques in the experimental part
of this paper.
\newline

{\it Incommensurate Transverse Interactions and MQT.}
Let us now consider the situation relevant to understanding the
physics of MQT in magnetic studies in which much smaller tunnel
splittings are probed ($\sim10^{-6}$ K). We will analyze the
behavior of the splitting for resonance $k=6$ ($m=-10$,$m'=4$)
versus the angle, $\phi$, of a small external transverse field.
The calculation has been done with the same $D$, $B$ and $C$
parameters of the Hamiltonian as those of the previous
calculations. We have used $|E|=E_{av}=3$ mK and a transverse
magnetic field, $H_T=0.35$ T, which corresponds to the situation
studied in the dc Landau-Zener relaxation experiments that will be
presented in section III.
\begin{figure}
\begin{center}\includegraphics[width=7cm]{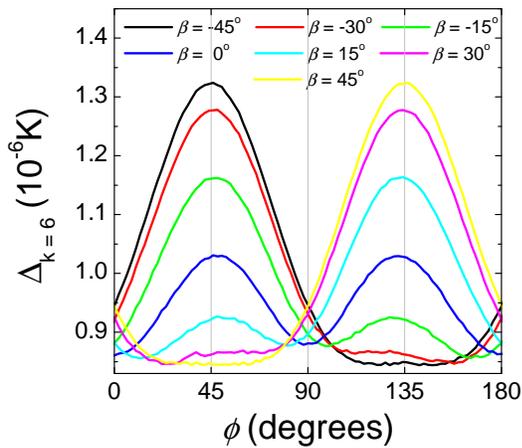}
\vspace{-2 mm}\caption{Ground state tunnel splitting,
$\Delta_{k=6}$, versus the orientation of a transverse field,
$H_T=0.35$ T, for $E=+3$ mK and different values of
$\beta$.}\vspace{-6 mm}
\end{center}
\end{figure}

The results are shown in fig. 9. This situation is substantially
different from that of the previous case of resonance $k=0$. The
tunnel splitting has a pattern that goes from two-fold maxima for
$\beta=\pm45^\circ$ to four-fold maxima for $\beta=0$.
Interestingly, {\it the maxima are in the directions of the medium
axes of $C$ independent of the direction of the medium axis of
$E$}. For example, for $\beta=-30^\circ$ the two maxima of the
tunnel splitting are at $\phi_{max}=45^\circ,135^\circ$ while the directions of
the medium
axis of $E$ are $\phi_{ME}=\beta+90^\circ=60^\circ, 240^\circ$. So,
for some $\beta$-values, the tunnel splitting exhibits a twofold
pattern of maxima with position determined by the fourth order
anisotropy. We have done these calculations for bigger values of
$E$ (not shown in this paper) that indicate that the range of
$\beta$-values around $\beta=0$ that exhibits fourfold symmetry is
narrower the bigger the value of $E$. However, for $E$-values
smaller than 30 mK the maxima positions are still determined by
the fourth order anisotropy. This constitutes an unexpected
result in a system with second and fourth order anisotropies that
is first pointed out in this work and has important consequences
for the interpretation of the experimental measurements that have
been previously published by some of the authors of this work
\cite{delBarco2,Hill2}, as will be explained in the experimental
sections III and IV.
\begin{figure}
\begin{center}\includegraphics[width=7cm]{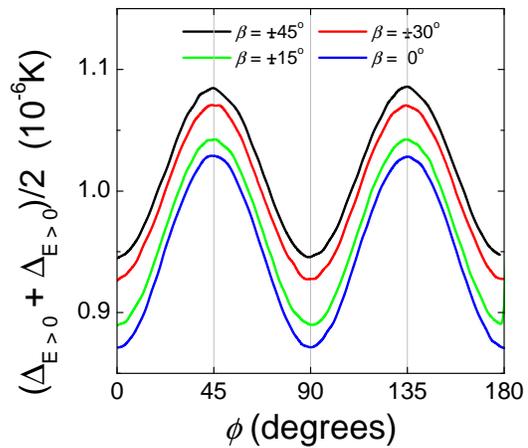}
\vspace{-2 mm}\caption{Average value of the ground state tunnel
splitting, $\Delta_{k=6}$, for opposite signs of $E$ ($|E=3|$ mK),
versus the orientation of a transverse field, $H_T=0.35$ T, and
for different values of $\beta$.}\vspace{-6 mm}
\end{center}
\end{figure}

We have calculated the average value of the tunnel splitting for
different angles $\beta$ assuming an equal population of molecules
with different signs of $E$. The results are shown in fig. 10. For
all values of $\beta$ there is clearly a fourfold rotation pattern of maxima
of the average tunnel splitting with symmetric maxima and minima
along directions determined by the fourth order anisotropy. Again,
the only method to determine the $E$ value and the relative
orientation between $C$ and $E$, $\beta$, is to study a subset of molecules
having only one sign of $E$. In
the latter case, depending on the value of $\beta$, there is a
possibility that the molecules selected will also show fourfold
symmetry (e.g. if $\beta\sim0$). It would then not be possible to
conclude that such molecules have a second order anisotropy. As we
will show in sections III and IV this is not the case
experimentally. We will show that a selection of a subset of
molecules with one sign of $E$ shows twofold symmetry which
indicates that the angle $\beta$ of misalignment is close to
$\beta\sim\pm45^\circ$ (specifically, $\beta=-30^\circ$).

{\it Disorder and Berry Phase Effects.} Quantum phase interference
is one of the most important phenomena observed in SMMs.
Interference effects (Berry phase) in MQT where first discovered
by Loss \cite{Loss2} and calculated for a nanomagnet with a
biaxial anisotropy by Garg \cite{Garg}. This phenomenon is due to
interference of the quantum tunneling trajectories of the
magnetization and has been clearly observed in two SMMs to date
\cite{Wernsdorfer,Wernsdorfer2}. The first observation of the
Berry phase was by Wernsdorfer and Sessoli \cite{Wernsdorfer} in
the Fe$_8$ SMM. Fe$_8$ has both second and fourth order transverse
anisotropy. The observation of quantum oscillations was done by
applying a transverse field along the direction of the hard axis
of the second order anisotropy (HE), which in this case also
corresponds to the direction of one of the hard axes of the fourth
anisotropy term, $\beta=0$. Fig. 11 shows the dependence of the
tunnel splitting of resonance $k=0$ on the external field applied
along $\phi=0$ (hard $E$ anisotropy axis) and $\phi=90^\circ$
(medium $E$ anisotropy axis). We have used the parameters given in
reference \cite{Wernsdorfer}.
\begin{figure}
\begin{center}\includegraphics[width=7cm]{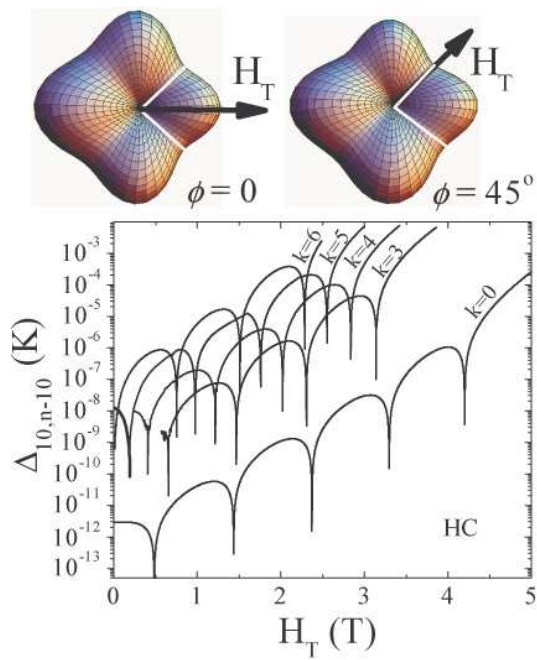}
\vspace{-2 mm}\caption{Fe$_8$ ground state tunnel splitting,
$\Delta_{k=0}$, versus the magnitude of a transverse field applied
along the hard (HE) and the medium (ME) anisotropy axes. The
parameters used in this calculation were taken from reference
\cite{Wernsdorfer}. The drawings above represent the $x-y$ plane
projections of the anisotropy barrier for Fe$_8$ in the presence
of a transverse field applied at different angles $\phi$. The
white lines show two hypothetical quantum tunneling trajectories.
When $H_T$ is applied along the hard anisotropy axis, $\phi=0$,
the barrier remains symmetric with respect to the field. In this
case, the trajectories interfere. For transverse fields not
aligned with the hard axis, an asymmetric distortion of the
barrier leads to non-equivalent MQT trajectories, destroying the
interference.}\vspace{-6 mm}
\end{center}
\end{figure}

The field spacing between quantum tunneling oscillations of fig.
11 can be described in terms of the anisotropy parameters $D$ and
$E$ of the Hamiltonian using a semiclassical approach \cite{Garg},
\begin{equation}
\label{eq.7} \Delta H=\frac{2k_B}{g\mu_B}\sqrt{2E(E+D)}\;,
\end{equation}
This is $\sim0.23$ T for Fe$_8$ which is smaller than the spacing
resulting from the exact diagonalization of the Hamiltonian
including fourth order transverse anisotropy, $\Delta H=0.41$ T,
and was observed in reference \cite{Wernsdorfer}. The reason is
that eq. (7) only considers the presence of a second order
anisotropy \cite{Pederson2}. Note that the authors of ref.
\cite{Wernsdorfer} assumed collinear anisotropies to fit the data,
thus the angle between the HE and HC1 axes is $\beta=0$ (see fig.
5). We also have used collinear second and fourth order anisotropy
terms in the calculated data shown in fig. 11. So we can see that
the effect of having a fourth order anisotropy that is collinear
with a second order anisotropy, $\beta=0$, only modifies the
pattern of oscillations but not their structure and shape. Quantum
tunneling oscillations were also observed in [Mn$_{12}$]$^{-2}$
which only has a second order transverse anisotropy
\cite{Wernsdorfer2}. In this case, the spacing between
oscillations was given by eq. 7.
\begin{figure}
\begin{center}\includegraphics[width=7cm]{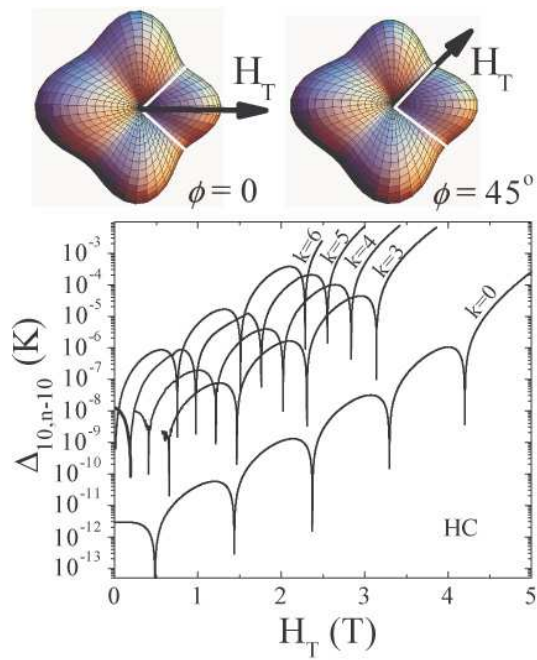}
\vspace{-2 mm}\caption{Transverse field dependence of the tunnel
splittings for Mn$_{12}$-ac , for several resonances, assuming
only fourth order anisotropy. The transverse field is applied
along one of the hard axes of $C$ (HC). The drawings above the
figure show the distortion of the anisotropy barrier due to a
transverse field applied at different angles with respect to the
hard axes of the fourth order anisotropy. White lines represent
different tunneling trajectories.}\vspace{-6 mm}
\end{center}
\end{figure}

Park and Garg \cite{Park2} calculated the quantum tunneling oscillations in a
system
with only fourth order transverse anisotropy using the Hamiltonian of
Mn$_{12}$-ac (Eq. (1)). Fig. 12 shows calculations of
the tunnel splitting for different resonances by using
eq. (1) with $D=556$ mK, $B=1.1$ mK and $C=0.03$ mK.
\begin{figure}
\begin{center}\includegraphics[width=7.5cm]{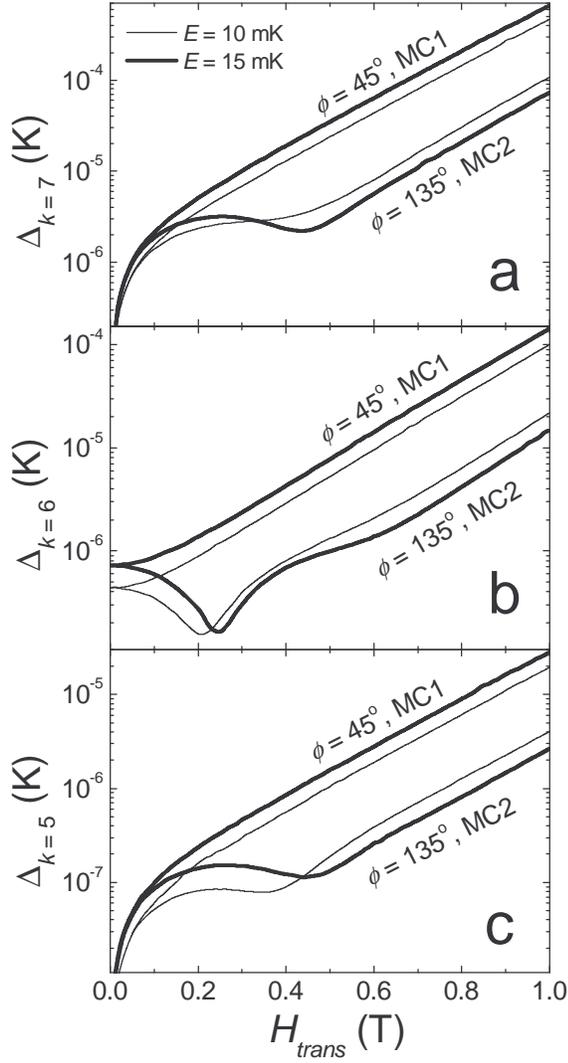}
\vspace{-2 mm}\caption{Transverse field dependence of the ground
state tunnel splitting of resonances $k=5$, $k=6$ and $k=7$ for
Mn$_{12}$-ac, with $E=10$ mK and $E=15$ mK. The transverse field
is applied along the $\phi=45^\circ$ and $\phi=135^\circ$
directions, which correspond to the positions of the maxima and
minima of $\Delta$ for a misalignment angle
$\beta=-30^\circ$.}\vspace{-6 mm}
\end{center}
\end{figure}

In order to take into account the effect of a misalignment,
$\beta\neq0$, between second and fourth order anisotropies we have
calculated the dependence of the ground state tunnel splitting on
the magnitude of an external transverse field applied along
different characteristic directions in the hard plane of a
molecule. For this, we have used $\beta=-30^\circ$, $D=548$ mK,
$B=1.17$ mK and $C=0.022$ mK and different values of $E>0$. For
clarity, this situation corresponds to having the following
directions for the characteristic transverse anisotropy axes:
$\phi_{HC}=0,90^\circ,180^\circ,270^\circ$,
$\phi_{MC}=45^\circ,135^\circ,215^\circ,305^\circ$,
$\phi_{HE}=\beta+n180^\circ=-30^\circ,150^\circ$ and
$\phi_{ME}=60^\circ,240^\circ$.

\begin{figure}
\begin{center}\includegraphics[width=8.5cm]{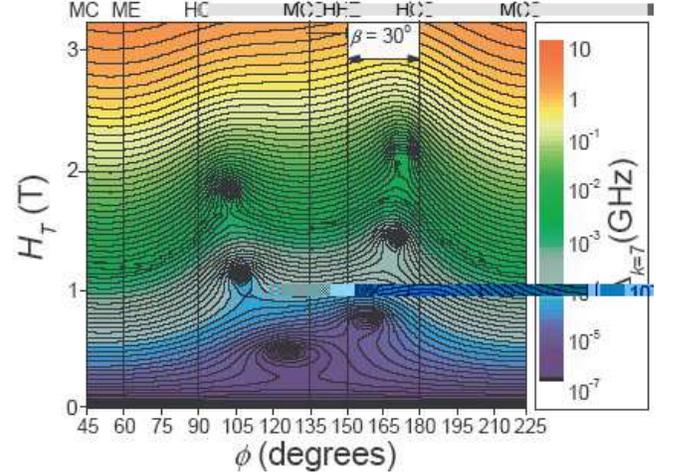}
\vspace{-2 mm}\caption{Color contour plot of the transverse field
dependence of the ground state tunnel splitting of resonance $k=7$
for different angles $\phi$ for a misalignment, $\beta=-30^\circ$,
between the hard anisotropy axes of $E$ and $C$. The vertical
lines represent the orientations of hard and medium axes of the
second and fourth order anisotropy terms. This misalignment
generates a new and interesting pattern of Berry phase zeros that
does not correspond to any of the anisotropies
separately.}\vspace{-6 mm}
\end{center}
\end{figure}
\begin{figure}
\begin{center}\includegraphics[width=8.0cm]{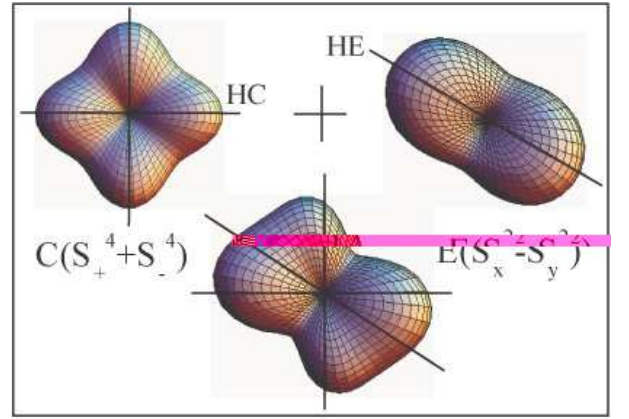}
\vspace{-2 mm}\caption{Projection of the anisotropy barrier onto
the $x-y$ plane for second (right-upper graphic) and fourth
(left-upper graphic) order transverse anisotropy terms. The
addition of both anisotropies leads to an asymmetric barrier
(lower graphic) that makes it difficult to see the orientations of
a transverse field that would generate equivalent quantum
tunneling trajectories and, therefore, Berry phase interference
phenomenon.}\vspace{-6 mm}
\end{center}
\end{figure}
Examining the result for $\beta=-30^\circ$ in fig. 9 (red curve)
one can see that the tunnel splitting has twofold symmetry with
maxima at $\phi=45^\circ,225^\circ$ and minima at
$\phi=135^\circ,315^\circ$. These directions correspond to the
medium axes of $C$ (MC1 and MC2). We have thus calculated the
dependence of the ground state tunnel splitting for resonances
$k=$ 5, 6 and 7 for the field applied along MC1 and MC2. These
are shown in fig. 13a, 13b and 13c respectively, for a transverse
field applied along $\phi=45^\circ$ (tunnel splitting maximum, MC1
axis) and $\phi=135^\circ$ (tunnel splitting minimum, MC2 axis)
using $E=10$ mK (thin lines) and $E=$ 15 mK (thick lines). Fig. 13
shows how the minimum in the tunnel splitting moves to higher
fields and becomes deeper as $E$ increases.

The results show a very different structure and shape of the Berry
phase oscillations as compared to the case when both
anisotropies are commensurate. Note that the orientations we used for
the applied transverse field correspond to the medium axes of the
fourth order anisotropy where one does not expect to have Berry
phase oscillations. Moreover, these two orientations,
$\phi=45^\circ$ and $\phi=135^\circ$, do not coincide with the
hard axis of the second order anisotropy, which for
$\beta=-30^\circ$ is along $\phi=-30^\circ+n180^\circ$.

In order to have a more complete picture of the effect of second and fourth
anisotropies with non-collinear axes
on the Berry phase phenomena we have calculated the dependence of
the ground state tunnel splitting of resonance $k=7$ on the
magnitude of a transverse field applied at different angles, from
$\phi=45^\circ$ (which correspond to one of the medium axes of
$C$) to $\phi=225^\circ$ (which correspond to the opposite
orientation of the field along the same hard $C$-axis). The
results are shown in a color contour plot in fig. 14. The first
thing to point out is that the
tunnel splitting still has zeros in this situation. However, the most
significant
fact is that these zeros do not appear at an angle characteristic of the
transverse anisotropies. Moreover, the structure, shape and
position of the zeros are completely independent of the Berry
phase corresponding to each anisotropy term separately. Note that
the calculations shown in fig. 13 correspond to transverse fields
applied along $\phi=45^\circ$ and $135^\circ$ which correspond to
to the directions of the maximum and minimum values of the tunnel
splitting of fig. 9 respectively. In the case of $\phi=135^\circ$
the tunnel splitting is close to one of the zeros ($\phi\sim
125^\circ$ and $H_T\sim 0.4$ T in fig. 14) but far from all the
others, explaining the observation of only one incomplete
oscillation in fig. 13a. This new and unusual structure of the
Berry phase zeros can be better understood by looking at the graphic
representation of the anisotropy barrier of fig. 15. The addition
of second (right-upper illustration in fig. 15) and fourth order (left-upper)
transverse anisotropies in a SMM leads to an asymmetric barrier
(center-lower) where symmetry does not permit a direct
identification of the field that generates equivalent quantum
tunneling trajectories that can interfere.

From the results, we can conclude that the combination of
incommensurate transverse anisotropy terms of different order in
the spin-operators can lead to an interesting situation in which
the resulting magnetic response does not depend in any simply way
on the form of either anisotropy term separately. The parameters
used in the above simulation were chosen because they are within
the range of values that can explain the experimental results
presented in sections III and IV.

\section{\label{sec:level1} Magnetic Relaxation Experiments}

We have carried out magnetic relaxation measurements in a single
crystal of deuterated Mn$_{12}$-ac in the pure quantum regime
($T=0.6$ K) in which relaxation is by MQT without thermal
activation over the anisotropy barrier \cite{Bokacheva}.
Deuterated crystals were studied because the purity of the
chemicals used in the synthesis leads to very high quality
crystals \cite{Hendrickson}.

We have used a high sensitivity micro-Hall effect magnetometer
\cite{Kent} to measure the magnetic response of a
Mn$_{12}$-ac single crystal of $\sim100$ micrometer size and
needle shape. We measure the longitudinal component of the
magnetization of the sample ($z$-component) by placing the crystal
with one of its faces parallel to the sensor plane and one end
just over the cross point of the micro-Hall sensor. The
magnetometer was placed inside a low temperature He$^3$ system. A
superconducting vector-field magnet was used to apply high
magnetic fields at arbitrary directions with respect to the crystal
axis.

\subsection{Landau-Zener Method}

The Landau-Zener (LZ) method has been used to study quantum
tunnel splittings in SMMs \cite{Wernsdorfer} and has become a
powerful tool to check for distributions of dipolar and nuclear
interactions \cite{Wernsdorfer2}, molecular micro-environments
\cite{Mertes,Hernandez2,delBarco,delBarco2} or internal transverse
magnetic fields \cite{delBarco3} in these materials. The method
consists in crossing a MQT resonance by sweeping the longitudinal
magnetic field at a constant rate, $\alpha=dH/dt$, and measuring
the fractional change of the magnetization in the process. The
anti-crossing of the spin levels $m$ and $m'$ of resonance
$k=m+m'$ is shown in the inset of fig. 16. For an ideal system of
non-interacting and monodisperse SMMs, and for low enough
temperatures (where thermal relaxation is negligible), the
normalized change of magnetization,
$(M_{before}-M_{after})/(M_{before}-M_{eq})$, is related to the
probability for a molecule to reverse its magnetization by quantum
tunneling. The bigger the MQT probability the larger the
magnetization change will be. This MQT probability is related to
the tunnel splitting, $\Delta$, by the LZ formula \cite{Zener},
\begin{equation}
\label{eq.8} {P_{LZ}=1-exp\left( -\frac {\pi\Delta^2}{2\nu_0}\frac
1\alpha\right)}\;\;,
\end{equation}
where $\nu_0=g\mu_B(2S-k)$ and $\nu_0\alpha$ is the energy sweep
rate. $R_{LZ}=1-P_{LZ}$ is the probability for a molecule to
remain in the metastable state $\mid m'\rangle$ after crossing the
resonance.

It is important to note that this relation is only valid if the
internal energy sweep rate a single molecule experiences is
proportional to the external sweep rate of the magnetic field.
This is not satisfied if there are internal dipolar or nuclear
fields. In fact, it has been shown in Fe$_8$ \cite{Wernsdorfer2}
and Mn$_{12}$-ac \cite{delBarco} that changing dipolar fields
lead to deviations from the LZ formula. In order to avoid these
effects the external sweep rate must be fast enough to have small
magnetization change (i.e. small changes of dipolar fields) in the
crossing process. The critical lower value of the magnetic field
sweep rate, $\alpha_c$, that is needed to avoid this situation has
been determined to be 10$^{-3}$ T/s for Mn$_{12}$-ac
\cite{delBarco}. Due to this, all the experiments presented in
this section have been conducted with $\alpha>\alpha_c$.
\begin{figure}
\begin{center}\includegraphics[width=7.5cm]{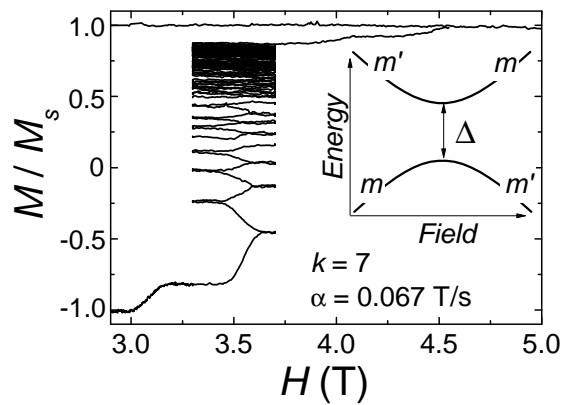}
\vspace{-2 mm}\caption{Landau-Zener multi-crossing experiment in a
Mn$_{12}$-ac single crystal measured at 0.6 K by sweeping the
longitudinal magnetic field at a constant rate,
$\alpha=6.6\times10^{-3}$ T/s, multiple times across the $k=7$
resonance. The inset shows a representation of the energy levels,
$m$ and $m'$, at the anti-crossing point.}\vspace{-6 mm}
\end{center}
\end{figure}

\subsection{Multi-crossing Landau-Zener Measurements}

When there is a distribution of quantum splittings in the sample
each molecule has a different MQT probability and the relaxation
of the magnetization should reflect this fact. In this case, the
MQT probability depends on the distribution of tunnel splittings
of the molecules that are in the metastable well before crossing a
resonance. After a crossing of a resonance, those molecules with
the largest tunnel splitting values and, consequently, the highest
MQT probability will represent the maximum contribution to the
fractional change of the magnetization. Correspondingly, those
molecules with smaller tunnel splitting values will remain in the
metastable well. Due to this, the LZ relaxation method can be used
to determine the distribution of tunnel splittings in a sample, and
to select different parts of the distribution for independent
study, as we will show in subsection III.C.

In order to extract the complete distribution of tunnel splittings
in Mn$_{12}$-ac we have used a modification of the LZ method
that consists of crossing a resonance multiple times, both for
increasing and decreasing fields. As we have stated previously, molecules in the
metastable well having the largest tunnel
splitting values are most likely to relax in any given crossing of a resonance.
Once a molecule has relaxed, it will no longer exhibit
dynamics in subsequent crossings of the same resonance. Thus, only those
molecules with the
largest probability of tunneling, and which did not already tunnel, can
contribute to the relaxation during subsequent crossings of the
resonance. The
repetition of this procedure many times enables a determination of
the distribution of tunnel splittings in the sample over several orders of
magnitude. We show an
example of this multi-crossing LZ procedure for resonance $k=7$ in
fig. 16.
\begin{figure}
\begin{center}\includegraphics[width=7.5cm]{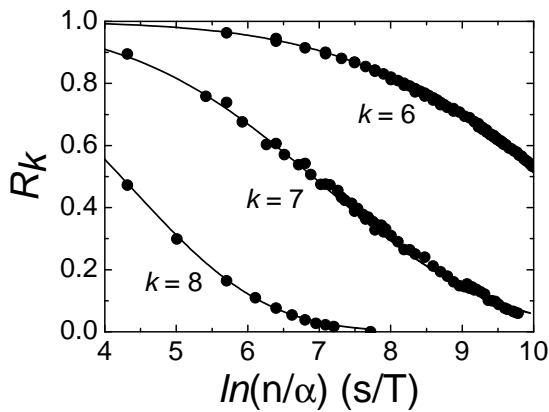}
\vspace{-2 mm}\caption{Landau-Zener probability to remain in the
metastable well in multi-crossing relaxations of resonances $k$ =
6, 7 and 8 using different sweep rates ($3.33\times10^{-3}$ to
$1.33\times10^{-2}$). All the measurements were performed
beginning with a saturation magnetization at $T$ = 0.6
K.}\vspace{-6 mm}
\end{center}
\end{figure}

In a multi-crossing LZ measurement the probability to remain in
the metastable well after crossing a resonance $n$-times is given
by,
\begin{equation}
\label{eq.9} {R_{LZn}=exp\left( -\frac {\pi\Delta^2}{2\nu_0}\frac
1\alpha_{eff}\right)}\;\;,
\end{equation}
where $\alpha_{eff}=\alpha/n$. If this expression describes the
physics, then relaxation curves recorded at different sweep rates
should scale when plotted as a function of the effective sweep
rate. This can be clearly observed in fig. 17, where we show LZ
multi-crossing relaxation measurements of resonances $k$ = 6, 7
and 8, carried out at different sweep rates $\alpha>\alpha_c$
($3.33\times10^{-3}$ to $1.33\times10^{-2}$). Small differences in
the results were observed in three different crystals that were
synthesized in the same way. These results clearly show that the
MQT relaxation rate is not exponential and indicate the presence
of a distribution of tunnel splittings within the sample. These
results confirm previous experimental observations of non
exponential relaxation in Mn$_{12}$-ac
\cite{Sangregorio,Perenboom,Mertes,Hernandez2}. Moreover, the
large fraction of the relaxation that we are able to observe with
this method gives direct information on the width of the
distribution of tunnel splittings.

We have assumed a log-normal distribution of tunnel splittings to
explain our observations. We take the form,
\begin{equation}
\label{eq.10} {f(x)=Aexp\left( -\frac
{(x-x_c)^2}{\sigma^2}\right)}\;\;,
\end{equation}
where $x=log\Delta$ and $x_c=log\Delta_c$; $\Delta_c$ and $\sigma$
represent the center and the width of the distribution, respectively.
The fit of the relaxation curves that are shown in fig. 17 (solid
lines) have been obtained by using,
\begin{equation}
\label{eq.11} {R(\alpha,n)=\int_{-\infty}^\infty
R_{lz}(\alpha,\Delta,n)f(log\Delta)dlog\Delta}\;\;,
\end{equation}
with $x_c$ and $\sigma$ as free parameters.
The resulting fits are in excellent accord with the
experiments. The resultant distribution functions for resonances
$k$ = 6, 7 and 8 are shown in fig. 18. The center of the
distribution increases with the resonance number, $k$, while the
width remains almost constant for different resonances, being
somewhat narrower for resonance $k$ = 8.
\begin{figure}
\begin{center}\includegraphics[width=7.5cm]{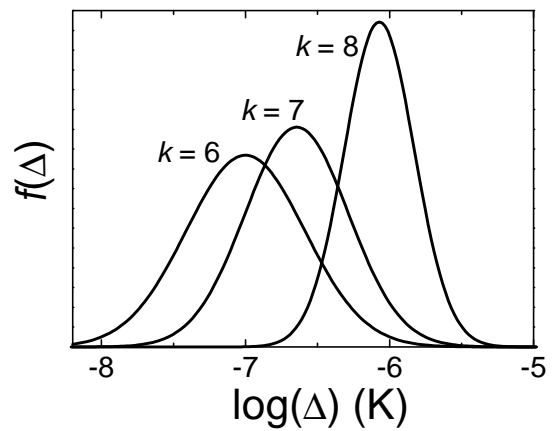}
\vspace{-2 mm}\caption{Log-normal distribution functions extracted
from the fit to the LZ relaxation curves of fig. 17.}\vspace{-6
mm}
\end{center}
\end{figure}

In section II.b we have discussed the two models that have been
proposed to explain the presence of a broad distribution of tunnel
splittings in terms of disorder. Before analyzing our results,
note that a second order transverse anisotropy
allows transitions for resonances $2i$, while a fourth order
anisotropy, which is imposed by the symmetry of the molecules,
only allows MQT relaxation for resonance numbers which are a multiple of 4,
$4i$. Thus, a comparison between relaxation curves recorded at
resonances $k$ = 6 and 8 should give us information about the
origin of the tunnel splittings in this material. The relation
between the tunnel splitting and the second order anisotropy is
given by the next formula which follows from perturbation theory
\cite{Garanin}:
\begin{equation}
\label{eq.12} {\ln\left(\Delta_k/g_k\right)/\xi_k=\ln\left(\frac
E{2D}\right)}\;\;,
\end{equation}
where $g_k$ and $\xi_k$ depend on $k$, $S$ and $D$ \cite{Garanin}
and were given in section II.a. Through this expression we can
infer the distribution of second order anisotropy parameters,
$f(ln(E/2D))$, by taking the log-normal distributions used to fit
the relaxation curves of resonances $k$ = 6 and $k$ = 8. The
results are shown in fig. 19. The fact that both distributions do
not scale when plotted as a function of $ln(E/2D)$ indicates that
second order anisotropy can not be the only origin of tunnel
splittings within the sample.
\begin{figure}
\begin{center}\includegraphics[width=7.5cm]{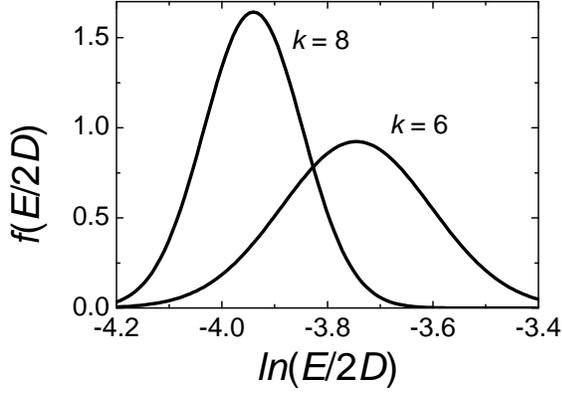}
\vspace{-2 mm}\caption{Distribution of the second order anisotropy
parameter inferred from the log-normal distribution functions of
resonances $k$ = 6 and 8 of fig. 18 using the expression of eq.
(11).}\vspace{-6 mm}
\end{center}
\end{figure}

The distribution function of the second order anisotropy parameter
predicted by the line dislocations model is given by
\cite{Chudnovsky2},
\begin{equation}
\label{eq.13} {f_L(x)\cong \frac
{1}{2\sqrt{\pi}\tilde{E}_{\tilde{c}}} exp\left(x-\frac {e^{2x}}{(2
{\tilde{E}_{\tilde{c}}})^2}\right)}\;\;,
\end{equation}
where $x\equiv\ln\tilde{E}$ with $\tilde{E}=E/2D$.
$\tilde{E}_{\tilde{c}}$ is the width of the distribution of the
anisotropy parameter $\tilde{E}$. $\tilde{E}_{\tilde{c}}$ depends
on the geometry of the crystal and on the concentration of
dislocations per unit cell, $c$. Note that the mean value and
width of the distribution given by this expression are not
independent variables. By using this distribution with eq. (11) we
have fit the relaxation curve recorded at resonance $k$ = 6.
The only free parameter in the fit is $\tilde{E}_{\tilde{c}}$. The
result is shown in fig. 20 (thin line) where the log-normal
distribution extracted from our previous fit of the same
relaxation curve has been included for comparison (thick line). We
have chosen the mean value to be at the same position as that of
the log-normal distribution. The value of $\tilde{E}_{\tilde{c}}$
used to fit the data corresponds to a concentration of
dislocations per unit cell of $c\sim10^{-4}$. Clearly, the width
of this distribution is many orders of magnitude bigger than the
log-normal distribution used to fit our data. This is due to the
fact that the distribution of $E$ predicted from the line
dislocations model has a most probable value at $E=0$, which
explains the long tail observed for low tunnel splitting values.
Consequently, line dislocations would produce a much broader
relaxation curve than that observed in the experiments. Our data
indicate that a distribution of second order anisotropy with a
non zero mode is needed to explain the MQT relaxation in
Mn$_{12}$-ac.

We want to note that we were able to fit the relaxation curves by
using a discrete multi-peak distribution of $E$ values similar to
that expected from the solvent disorder model \cite{Cornia}.
However, it was necessary to include a Gaussian width to each peak
of the distribution in order to fit of the data \cite{delBarco}.
The values of the peak centers of this distribution for resonance
$k$ = 6 are $x_{c,1}=$-7.19 (-7.0525), $x_{c,2}=$-8.55 (-8.1749),
and $x_{c,3}=$-6.60 (-6.7995), (the values in parenthesis are
extracted from ref. \cite{Cornia}). The width of each peak is
$W_i=x_{c,i}$/50 and the height of each peak was taken to be
proportional to the population of the corresponding isomer given
in ref. \cite{Cornia}.
\begin{figure}
\begin{center}\includegraphics[width=7.5cm]{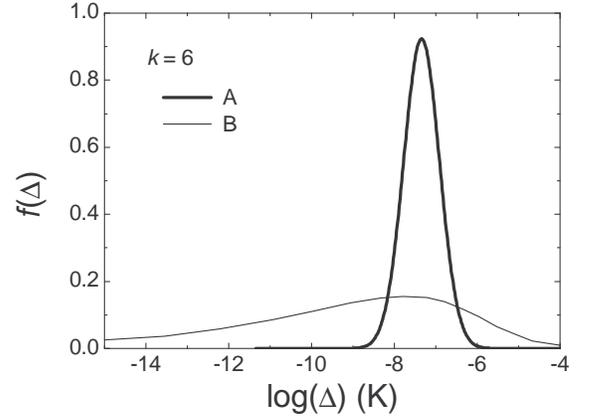}
\vspace{-2 mm}\caption{Tunnel splitting distribution of resonance
$k$ = 6 expected from the line-dislocation model (thin line)
compared with the log-normal distribution that fits the
experimental data (thick line).}\vspace{-6 mm}
\end{center}
\end{figure}

To conclude this subsection, we have shown that a multi-crossing
LZ method allows the determination of the complete distribution of
tunnel splittings for several resonances in SMMs. Our results
suggest that a distribution of second order transverse
anisotropies with a non zero mode is necessary to explain
the experimental data. The solvent disorder model provides such a
source and a discrete multi-peak distribution of tunnel
splittings can be used to fit our data. However, an additional source
of disorder (i.e. line-dislocations or point defects) that
introduces a small broadening of the these peaks is also necessary
to model the experimental data.

\subsection{MQT Symmetry Measurements}

In this subsection we will present LZ relaxation experiments
carried out in the pure quantum regime ($T$ = 0.6 K). In order to
check the MQT symmetry imposed by the transverse terms of the
Hamiltonian we have studied the LZ relaxation of the magnetization
by sweeping an external longitudinal field, $H_L$, at a constant
rate, $\alpha$, across a resonance $k$ in the presence of an
external transverse field, $H_T$, applied at arbitrary directions,
$\phi$, with respect to the crystallographic axes of a
Mn$_{12}$-ac single crystal. As we have shown in section II
(i.e. see fig. 4), MQT has an oscillatory response as a function
of the orientation of a transverse field. This leads to maxima and
minima in the MQT relaxation rates whose positions and symmetry
depend on the transverse anisotropy term that generates the tunnel
splittings. To recall, a fourth order anisotropy term would
generate a fourfold rotation pattern in the MQT probability with
maxima spaced by 90 degrees, while a twofold rotation pattern with
spacing between maxima of 180 degrees is expected from a second
order anisotropy term in the Hamiltonian. When two non-collinear
different order transverse anisotropy terms are present, the
symmetry of the MQT relaxation rates depends on the relative
orientation between the anisotropy axes, as discussed in Sec. IIB.

A single crystal of Mn$_{12}$-ac SMMs was placed over a high
sensitivity micro-Hall magnetometer as described in the first
paragraph of this section. However, for these studies it is very
important to know the exact orientation of the crystallographic
axes with respect to the direction of the external magnetic field.
Fig. 21 shows a sketch of the orientation of the $c$-axis of the
crystal with respect to the external magnetic field. Note that, in
Mn$_{12}$-ac, the $c$-axis corresponds to the easy magnetic
axis of the molecules. However, there is a misalignment of
$\varphi\sim12^\circ$ between the transverse magnetic axes
(imposed by the fourth order anisotropy of the molecules) and the
crystallographic axes. This is also shown in fig. 21.
\begin{figure}
\begin{center}\includegraphics[width=7.5cm]{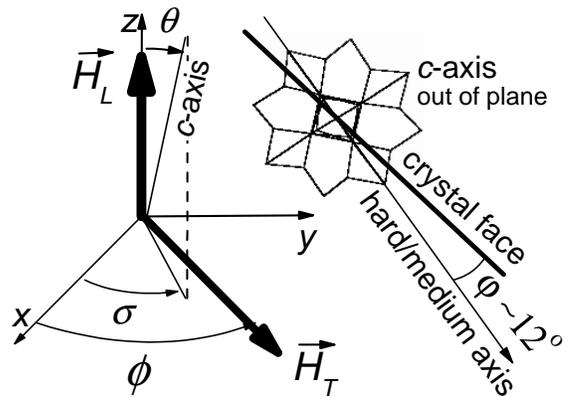}
\vspace{-2 mm}\caption{Schematic representation of the orientation
of the $c$-axis (easy axis) of a crystal with respect to the axes
of the applied external magnetic field ($x$,$y$,$z$). The
misalignment is determined by $\theta$ and $\sigma$. The
misalignment between the transverse magnetic axes of the
Mn$_{12}$-ac molecules, and one of the faces of the crystal
($\varphi\sim12^\circ$), is also shown.}\vspace{-6 mm}
\end{center}
\end{figure}

During the manual alignment of the crystal one of its faces was
placed coplanar with the micro-Hall sensor plane with the help of
a microscope. However, there exists an uncertainty of about $\pm$5
degrees in this orientation. This misalignment is represented by
the angles $\theta$ and $\sigma$ in fig. 21 and can be determined
experimentally through the magnetic measurements described below. The
main implication of this misalignment is that there is a
transverse field component due to the high longitudinal field
applied along the $z$-axis that depends on the angles $\theta$,
$\sigma$ and $\phi$. The latter is the angle of application of the
external transverse field with respect to the $x$-axis. Due to
this, the total transverse field felt by the molecules in the
presence of a constant transverse field, $H_T$, applied along
$\phi$ and with a longitudinal field, $H_L=H_zcos\theta\sim H_z$
($cos\theta\sim 1$ for $\theta$ small) is given by,
\begin{equation}
\label{eq.14} {H^2\sim
[H_Tcos(\phi-\sigma)-H_Lsin\theta]^2+[H_Tsin(\phi-\sigma)]^2}
\end{equation}
Thus, the transverse field felt by the molecules will have a
minimum value for $\phi=\sigma$ and a maximum for
$\phi=\sigma+180^\circ$ with $H _L>0$. The opposite situation
would be found for $H_L<0$. As we will show, once the misalignment
angles are known, an algorithm can be used to correct the applied
fields in order to have a constant transverse field during the
measurements, independent of the angle of application of the
external fields.

In our first experiment we have studied the MQT relaxation rates
of several resonances in the presence of a constant transverse
field applied at arbitrary directions with respect to the
crystallographic axes. The experiment was done as follows: We
start with the initial magnetization of the sample equal to
positive saturation, $M_{initial}=+M_s$, by applying a high
longitudinal magnetic field, $H_L=6T$. For this initial situation,
all of the molecules of the crystal were in the $m=+10$ level in one
of the energy wells. Then we turn on a transverse magnetic field,
$H_T$ = 0.4 T, applied along a direction, $\phi$, and sweep the
longitudinal field at a constant rate, $\alpha=6.6\times10^{-3}$
T/s. We measured the magnetization change in several resonances
$k$ and determined the MQT probability,
$P_{LZ}=(M_{before}-M_{after})/(M_{before}-M_{eq})$, where, in
this case, $M_{eq}$ = $-M_s$. Note here that all the molecules
within the crystal contribute to the relaxation. We repeated this
procedure for different angles $\phi$ from 0 to 360 degrees.
\begin{figure}
\begin{center}\includegraphics[width=7.5cm]{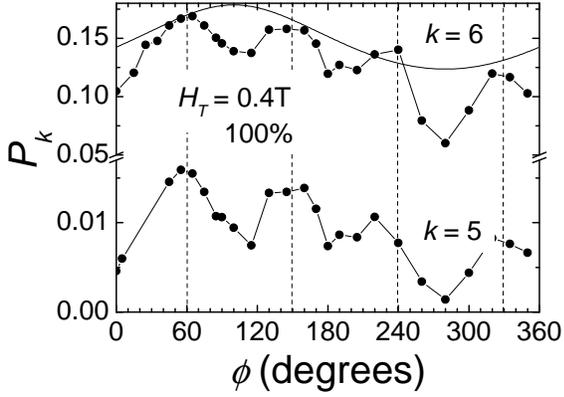}
\vspace{-2 mm}\caption{Measured MQT relaxation probability for
resonances $k$ = 6 and $k=7$, as a function of the orientation of
the applied transverse field, $H_T$ = 0.4 T, relative to one of
the faces of the crystal. The measurements where carried out
starting from saturation (all the molecules within the crystal
contributed to the relaxation).}\vspace{-6 mm}
\end{center}
\end{figure}

The behavior of the measured MQT probability as a function of
$\phi$ is shown in fig. 22 for resonances $k$ = 5 and $k$ = 6. The
results clearly show fourfold maxima in the tunneling probability spaced by 90
degrees
($\phi_{max}=60^\circ,150^\circ,240^\circ$ and $330^\circ$) for
both resonances. Note that $k$ = 5 and 6 are the first observed
resonances and $\sim35\%$ of the magnetization relaxes. This means
that molecules that contribute to this relaxation are mainly those
with among the biggest tunnel splitting values within the
distribution.

There is also a one-fold contribution that is represented by a
continuous line in fig. 22 with the result for resonance $k$ = 6
\cite{note1}. This oscillation is due to the misalignment of the
$c$-axis of the crystal with respect to the applied magnetic field.
This misalignment is represented by the angles $\theta$ and
$\sigma$ and can be determined through magnetic measurements. From
the results shown in fig. 22 we find $\sigma=80^\circ$
(where we observe the maximum value of the one-fold contribution).
To obtain $\theta$, we measured the behavior of the MQT
probability of resonance $k$ = 6 for different magnitudes of the
transverse field from -0.4 to 0.4 T applied along $\phi=\sigma$.
The transverse field value for which the probability is minimum
(null real transverse field) gives the angle of misalignment
$\theta$ through $H_T(P_min)=H_Lsin\theta$, where $H_Lsin\theta$
is the transverse projection of the longitudinal field. The value
extracted for the misalignment is $\theta=0.3^\circ$. The effect
of this misalignment on the transverse field is larger the greater
the resonance number. We do not show the results for higher
resonances, such as $k$ = 8, because the fourfold symmetry is
almost unobservable due to the high one-fold contribution of the
misalignment.

\begin{figure}
\begin{center}\includegraphics[width=7.5cm]{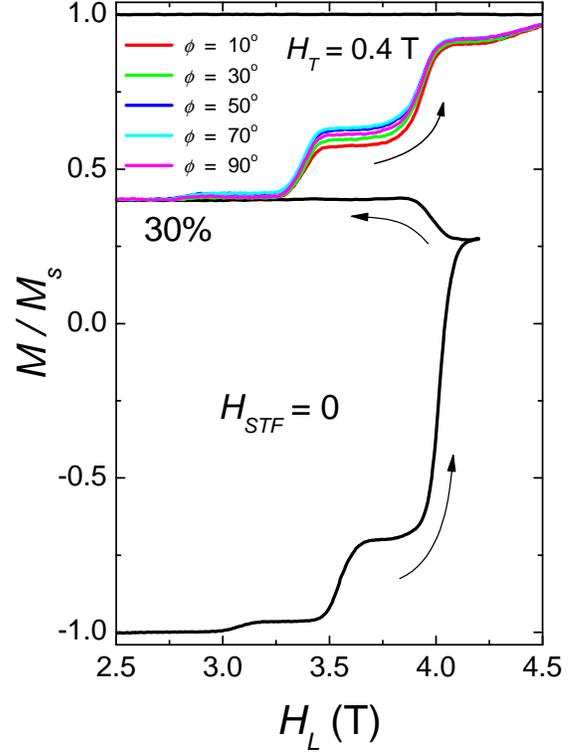}
\vspace{-2 mm}\caption{Selection of 30$\%$ of the molecules with
the smallest tunnel splittings within the distribution. Starting
at $M=-M_s$, the longitudinal field is swept from 0 to 4.2 T, then
back to 0 in the absence of a transverse field (black line). After
this procedure, only those molecules that have not relaxed
(30$\%$) remain in the metastable well. A transverse field of 0.4
T is then applied at an angle $\phi$, and the longitudinal field
is swept again to a high positive value. The process is repeated
for different $\phi$ angles (lines with different
colors).}\vspace{-6 mm}
\end{center}
\end{figure}
In order to measure the response of the MQT probability versus the
angle $\phi$ in other parts of the tunnel splitting distribution
we have conducted LZ relaxation experiments in the following
manner. We select a small fraction of molecules with the smallest
tunnel splittings of the distribution, in contrast to the biggest
values that were analyzed in the previous experiments. The
measurement method is presented in fig. 23. We start with $M=-M_s$
by applying a high negative longitudinal magnetic field, $H_L$ =
-5 T. Then we sweep the longitudinal field at a constant rate,
$\alpha=6.6\times 10^{-3}$ T/s, up to $H_L=+$ 4.2 T (just after
crossing resonance $k$ = 8) and sweep it back to zero, crossing
again resonances $k$ = 8, 7, 6... The whole selection process is
done in the absence of a transverse field. After that, the final
magnetization of the sample is $M=0.4M_s$. This means that 70$\%$
of the molecules have relaxed to the stable well or, in other
words, only 30$\%$ of the molecules have remained in the
metastable well and will contribute to further relaxation. The
latter are those molecules with the smallest tunnel splitting
values within the distribution. After this, we turn on a
transverse field, $H_T$ = 0.4 T, applied at an arbitrary angle,
$\phi$, with respect to the crystallographic axes of the sample.
Then we sweep the longitudinal field to a high positive value,
crossing the resonances again. We repeated this procedure for
different orientations of the transverse field from 0 to 360
degrees.
\begin{figure}
\begin{center}\includegraphics[width=7.5cm]{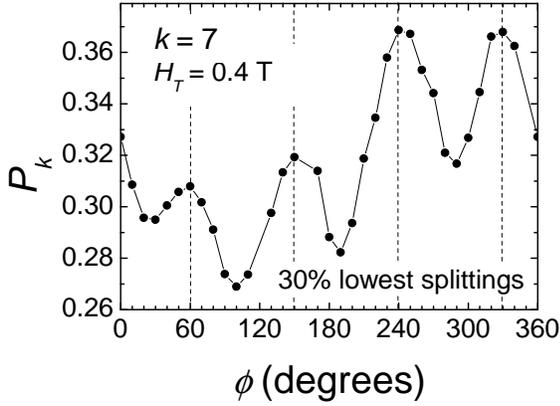}
\vspace{-2 mm}\caption{MQT probability for resonance $k=7$ versus
the orientation of the applied transverse field, $H_T$ = 0.4 T,
relative to one of the faces of the crystal. A previous selection
process was used to study only 30$\%$ of the molecules with the
smallest tunnel splitting values within the distribution. The
results were extracted from longitudinal magnetic field relaxation
curves recorded at a constant sweep rate, $\alpha=6.6\times
10^{-3}$ T/s.}\vspace{-6 mm}
\end{center}
\end{figure}

The MQT probability of resonance $k=6$ is shown as a function of
$\phi$ in fig. 24. The results show the same fourfold symmetry
pattern with maxima placed at the same positions as those in the
experiment shown in fig. 22. As these two experiments study the
relaxation of two different parts of the distribution of tunnel
splittings (low and high ends of the distribution), we can conclude
that the fourfold symmetry of MQT is a property of a significant
fraction of the molecules within the crystal.

In principle, the four fourfold rotation pattern is consistent
with a fourth order transverse anisotropy term, $C(S_+^4+S_-^4)$,
in the spin-Hamiltonian (see fig. 4b). For positive $C$, the four
maxima generated by this term should be at
$\phi_{max}=45^\circ,135^\circ,225^\circ$ and $315^\circ$. There
is a difference of 15 degrees between these values and those
observed in our experiments,
$\phi_{max}=60^\circ,150^\circ,240^\circ$ and $330^\circ$. This is
due to the misalignment, $\varphi=12^\circ$, between the hard
anisotropy axis of the molecules and one of the faces of the
crystal, which we use as the origin of our $\phi$ rotation. There
still is a difference of 3 degrees that is within the accuracy
with which we orient the crystal ($\pm$5 degrees). However, the
value of $C\sim 3\times 10^{-5}$ K cannot explain the difference
between the maximal and minimal magnitudes of the measured MQT
probability. The observed normalized changes,
$(P_{max}-P_{min})/P_{max}$, in the experiment with 100$\%$ of
the molecules contributing to the relaxation are $\sim$0.9 for
$k=5$ and $\sim$0.6 for $k=6$ (see fig. 22). Whereas, with this
value of $C$, we expect this change to be within the noise of the
measurement.

In order to determine whether this symmetry is intrinsic to the
Mn$_{12}$-ac molecules and its origin, we have carried out
experiments designed to select different parts of the distribution
by using transverse fields in the selection process. Note that, in
the previous experiment, we selected the SMMs with the smallest tunnel
splittings in the absence of transverse fields. Now, we use a
selection transverse field (STF), $H_{STF}$, applied at an angle,
$\phi_{STF}$, during the preparation of the initial state of the
system. In this case, those molecules with the medium anisotropy
axis aligned with the STF have larger tunnel splitting values
(larger relaxation probability) and can be selected for further
study.
\begin{figure}
\begin{center}\includegraphics[width=7.5cm]{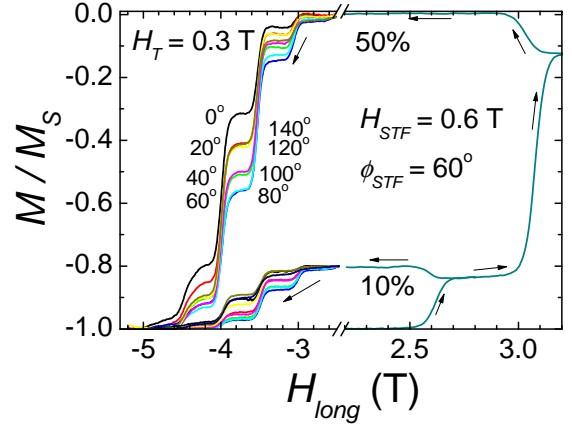}
\vspace{-2 mm}\caption{Selection processes carried out by sweeping
a positive longitudinal field across resonances $k=5$ and $k=6$,
in the presence of a selection transverse field, $H_{STF}$ = 0.6
T, applied at a selection angle $\phi_{STF}=60^\circ$. Both $10\%$
and $50\%$ of the molecules with the largest tunnel splitting
values were selected in separate processes for this experiment.
These populations were then studied on the negative side of the
longitudinal field hysteresis curve, in the presence of a
transverse field, $H_T$ = 0.3 T, applied at different angles
$\phi$ from 0 to 360 degrees.}\vspace{-6 mm}
\end{center}
\end{figure}

The measurements with this selection process are shown in fig. 25.
First we apply a high longitudinal field to saturate the
magnetization of the system and sweep the field back to zero,
having at the end a magnetization, $M=-M_s$. Then we turn on a
selection transverse field, $H_{STF}$ = 0.6 T, applied at the
angle $\phi_{STF}=60^\circ$, where one of the four maxima were
observed in the experiments carried out with the whole sample, and
we sweep the longitudinal field at a constant rate to a positive
value and sweep back to zero. The value of this field is chosen
depending on how much relaxation we want in the selection process:
for a selection of 50$\%$ of the biggest splittings, we sweep the
field up to 3.2 T allowing the system to relax in resonances $k$ =
5 and 6; for a selection of the 10$\%$ of the biggest splittings,
we only allow the system to relax in resonance $k=5$. The final
states of the magnetization are $M=0$ and $M=-0.8M_s$
respectively. After the selection process we sweep the
longitudinal field down to -5.5 T at a constant rate,
$\alpha=6.6\times 10^{-3}$, in the presence of a transverse field,
$H_T$ = 0.3 T, applied at difference angles $\phi$ with respect to
one of the faces of the crystal. We repeated this procedure for
different angles $\phi$ from 0 to 360 degrees. In a separate
experiment, we repeated the selection of 50$\%$ of the molecules
with the largest splittings within the distribution by applying a
selection transverse field, $H_{STF}$, along the angle in which a
complementary maximum was observed, $\phi_{STF}=150^\circ$.
\begin{figure}
\begin{center}\includegraphics[width=7.0cm]{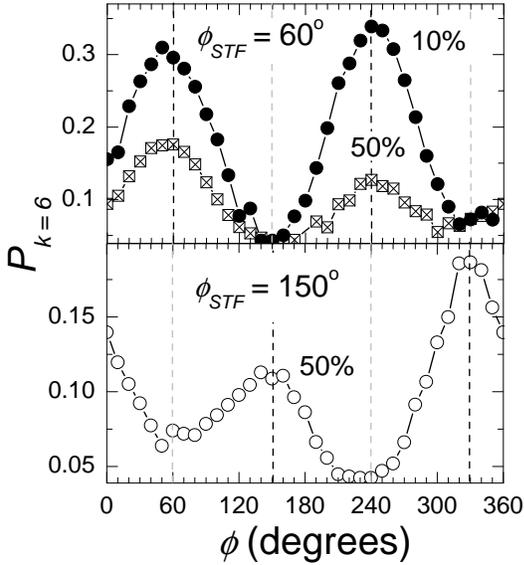}
\vspace{-2 mm}\caption{Behavior of the MQT probability versus the
orientation of a transverse field, $H_T$ = 0.3 T. The molecules
which relax were previously selected from the tunnel splitting
distribution using a selection transverse field, $H_{STF}$,
applied along the directions where the two complementary maxima
where observed in the experiments on the whole sample (fig. 24),
i.e. $\phi_{STF}=60^\circ$ (upper figure, for both 10$\%$ and
50$\%$ of the largest splittings in the distribution), and
$\phi_{STF}=150^\circ$ (lower figure, for 50$\%$ of the largest
splittings).}\vspace{-6 mm}
\end{center}
\end{figure}

The results are shown in fig. 26. In all
selections the MQT probability shows a twofold rotation
pattern with maxima spaced by 180 degrees. For the selection in
which $H_{STF}$ is applied along $\phi_{STF}=60^\circ$, the two
maxima are at $\phi_{max,1}=60^\circ$ and
$\phi_{max,2}=240^\circ$, for both fractions of molecules selected
(10$\%$ and 50$\%$). Moreover, when the selection field is applied
along the position of a complementary fourfold maxima,
$\phi_{STF}=150^\circ$, the twofold maxima,
$\phi_{max,1}=150^\circ$ and $\phi_{max,2}=330^\circ$, are
displaced by 90 degrees with respect to the previous case.

The observation of a twofold rotation pattern in the MQT probability is
clear evidence of a second order transverse anisotropy lower
than that imposed by the site symmetry of the molecule (four-fold). This is in
excellent agreement with the solvent disorder
model proposed by Cornia et al. \cite{Cornia} where there is an
expectation of equal populations of molecules with opposite signs
of the second order anisotropy. A considerable increase of the
change of probability between maxima and minima is also observed
as the fraction of molecules with highest tunnel splitting values
becomes smaller, indicating the fact that the molecules with
largest splitting values have bigger values of the second order
anisotropy.
\begin{figure}
\begin{center}\includegraphics[width=8.6cm]{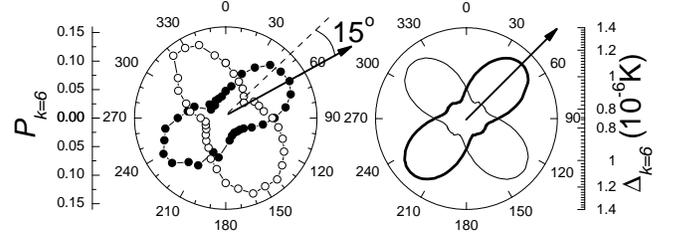}
\vspace{-2 mm}\caption{{\bf Left:} Experimentally determined angle
dependence of the MQT probability for resonance $k=6$ for the two
complementary directions of the transverse selection field. The
one-fold contribution arising from the misalignment angles
$\theta$ and $\sigma$ has been corrected for clarity. {\bf Right:}
Calculated tunnel splitting for resonance $k=6$, $\Delta_{k=6}$,
for $\beta=-30^\circ$ and opposite signs of $E$; the calculations
assume the same values of $H_T$, $D$, $E$ and $C$ as those in fig.
9.}\vspace{-6 mm}
\end{center}
\end{figure}

In section II we stated that incommensurate anisotropy terms
would considerably modify the magnetic response of the molecules
to the orientation of a transverse magnetic field, depending on the
angle of misalignment, $\beta$, between both anisotropies (see
eqs. (5) and (6) and fig. 5). The symmetry of the MQT probability
expected from this model depends on several parameters like the
resonance number, $k$, and/or the misalignment angle, $\beta$.
It turns out from this model
that for $|E|<$ 30mK (bigger than the maximum $E$ value expected
from the solvent disorder model) the symmetry of the MQT
probability mainly depends on the angle of misalignment, $\beta$,
going from fourfold for small $\beta$ values to twofold for big
$\beta$ values. Taking a given $E$ value, the transition between
fourfold to twofold maxima patterns as a function of $\beta$
depends on the resonance number $k$. For
example, for resonance $k=0$ (fig. 8A), the fourfold maxima
pattern is observed for every misalignment value, where a slight
twofold modulation is due to the second order anisotropy. For
bigger resonances (i.e. $k=6$), the transition between these two
MQT probability symmetries is cleaner (see fig. 9). For
misalignment angles, $|\beta|<20^\circ$, the MQT probability shows
fourfold symmetry modulated by the second order anisotropy.
However, for angles $|\beta|>20^\circ$, the MQT probability
symmetry is twofold. Consequently, the observation of twofold
symmetry in our experiments shows that the angle of misalignment
between both anisotropy terms is greater than 20 degrees. In
section IV we will present high frequency EPR experiments that
show that the angle of this misalignment is $\beta=-30^\circ$. In
fig. 27 we show a comparison between the experimental observation
of resonance $k=6$ for both transverse selection fields with
50$\%$ of the biggest splittings (left polar plot) and the
calculated splitting (right polar plot) corresponding to resonance
$k=6$ and $\beta=-30^\circ$ with opposite signs of $|E|=3 mK$. The
difference of $\sim$15 degrees between the experimental results
and the calculations is also shown in this figure. As we said
before, this difference is due to the fact that we measure the
angle $\phi$ with respect to one of the faces of the crystal while
the transverse magnetic axes of the molecules are rotated from the
faces of the crystal by $\sim$12 degrees.

An estimation of the values of $E$ needed to explain the
experimental observations of the oscillation of the MQT
probability presented in this subsection are: a) $E\sim$ 0.5 mK
for the result corresponding to the 30$\%$ of the smallest
splittings of the distribution (fig. 24), b) $E\sim$ 2.5 mK for
the result with the whole distribution (fig. 22, upper curve) and
with 50$\%$ of the biggest splitting in the selection with
$H_{STF}=60^\circ$ and 50$^\circ$ (fig. 26), and c) $E\sim$ 10 mK
for the result corresponding to 10$\%$ of the biggest splittings
of the distribution (fig. 26, solid circles). These values are in
excellent agreement with the results obtained by high frequency
EPR experiments (presented in section IV) and by recent
density functional theory calculations \cite{Pederson}. However,
they are slightly larger than the values initially proposed in
ref. \cite{Cornia}.

\subsection{Berry Phase Measurements}

The ability to select a subset of molecules with a narrow
distribution of tunnel splittings and, for example, different
signs of the second order transverse anisotropy allows us to study
the behavior of the MQT relaxation as a function of a transverse
field. As we anticipated in section II, quantum phase interference
(Berry phase) would lead to zeros of the tunnel splittings (i.e.
absence of magnetic relaxation) for several values of a transverse
field applied along the hard anisotropy axis of the molecules, and
we discussed how the pattern of the oscillations can be modified
by the presence of two non-collinear transverse anisotropies.
\begin{figure}
\begin{center}\includegraphics[width=8.6cm]{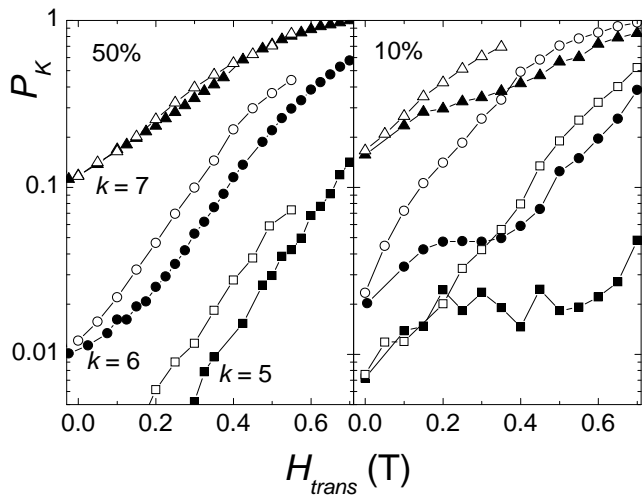}
\vspace{-2 mm}\caption{MQT probability for resonances $k=5$, 6 and
7 as a function of the magnitude of a transverse field applied
along $\phi=60^\circ$ (open symbols) and $\phi=150^\circ$ (solid
symbols). The initial state of the sample was prepared by
selection of 50$\%$ (left) and 10$\%$ (right) of the molecules
with the largest tunnel splitting values, using a selection
transverse field, $H_{STF}$ = 0.6 T, applied along
$\phi_{STF}=60^\circ$. The two transverse field orientations
correspond to the first maximum ($\phi=60^\circ$) and minimum
($\phi=150^\circ$) of the twofold rotation pattern.}\vspace{-6 mm}
\end{center}
\end{figure}

For these experiments we have used the same preparations of the
initial states of the system as those shown in fig. 25. These
initial states correspond to a selection of $50\%$
and $10\%$ of the molecules having the larger tunnel splitting values
within the distribution. The selection of both
initial states is done by applying a selection transverse field,
$H_{STF}$ = 0.6 T, at an angle $\phi_{STF}=60^\circ$. As we have
shown, this procedure mainly selects those molecules with one
$E$-sign, as is observed in the twofold transverse field rotation
pattern of the MQT probability of fig. 26, with maxima at
$\phi_{max}=60^\circ,240^\circ$ and minima at
$\phi_{max}=150^\circ,330^\circ$. To study the behavior of the MQT
probability as a function of the magnitude of the applied
transverse field we apply a transverse field, $H_T$, along the
direction of the first maximum, $\phi=60^\circ$, after the
selection process. We then sweep the longitudinal field at a
constant rate to $-5.5$ T, measuring the change of magnetization in
each resonance crossing. We follow the same procedure for different
values of the transverse field from $H_T=0$ to 0.7 T. Moreover, we
repeated the same measurement by applying the transverse field
along the direction of the first minima $\phi=150^\circ$
\cite{note2}. The results of the MQT probability for resonances $k$ =
5, 6 and 7 are shown in fig. 28. In the left graph of the figure
are the results obtained with 50$\%$ of the molecules with the largest
tunnel splitting values. In the right graph we show the results for
$10\%$ of the biggest splittings. The measurements with the
transverse field applied along the first of the twofold maxima,
$\phi=60^\circ$, are represented by open symbols, while the solid
symbols correspond to measurements with the transverse field
applied along the first of the minima, $\phi=150^\circ$.

There are several important aspects to this figure. a) There is
vertical shift between the curves corresponding to different
directions of application of the transverse field. The shift is
bigger in the case of the selection of the $10\%$ of the biggest
splittings within the distribution. This is consistent with the
observations of fig. 26, and supports the assumption of a
distribution in the magnitude of the second order anisotropy. b)
MQT probability increases exponentially with the magnitude of the
transverse field. This is expected from the exponential dependence
of the LZ probability on the tunnel splitting shown in eq. (8),
and the power law dependence of the tunnel splitting on the
magnitude of the transverse field (see eq. (3)) \cite{Garanin}.
There are significant deviations from the exponential behavior in
the right hand graphic for a transverse field applied along the
first of the twofold minima, $\phi=150^\circ$. The largest
deviations are observed at fields $H_{p}(k=5)\sim$ 0.45 T,
$H_{p}(k=6)\sim$ 0.3 T and $H_{p}(k=7)\sim$ 0.35 T. This is
reminiscent of the Berry phase observed in Fe$_8$
\cite{Wernsdorfer}.

The results shown on the right graph of fig. 28 can be compared with
the calculations of the tunnel splitting versus the transverse
field shown in fig. 13 that were performed by taking into account two
incommensurate anisotropy terms in the Hamiltonian (eqs. (5) and
(6)). In fact, the values, $E=10-15$ mK, used in these
calculations were chosen according to the results shown in fig. 28
and agree with the values extracted from high frequency EPR
measurements (section IV). The agreement between theory and
experiments is very good and constitutes the first evidence of
quantum interference phenomena in a SMM system with incommensurate
transverse anisotropies.

\subsection{Summary of magnetic relaxation experiments}

We have shown in this section that LZ magnetic relaxation
experiments allow us to determine the complete distribution of
tunnel splittings in Mn$_{12}$-ac. The results obtained
through a multi-crossing LZ method show that a distribution of
second order anisotropies with non-zero mode is required in order
to explain our data, such as in the solvent disorder model
proposed by Cornia et al. \cite{Cornia}. LZ relaxation experiments
carried out in the presence of a transverse field applied at
arbitrary directions with respect to the crystallographic axes of
the sample enabled studies of the symmetry of the MQT probability.
We have shown that the MQT probability has a general fourfold rotation
pattern as a function of the orientation of a transverse field.
This is associated with equal populations of molecules with
opposite signs of a second order transverse anisotropy. The LZ
method allows the selection of a subset of molecules with
different values of the tunnel splitting for further study. By applying a
transverse field in the selection process, we can select a fraction of molecules
in the sample with lower
symmetry and with different signs of $E$. Using this selection
procedure we have studied a small fraction of molecules with one
sign of $E$ and with the largest tunnel splitting values within the
distribution. These show an unusual Berry phase phenomena for
several transverse field values that does not lead to complete
zeroes in the tunnel splitting. Our results on the symmetry of MQT
can be explained in terms of incommensurate
transverse anisotropies in the Hamiltonian that
explain, among other things, why the observed Berry phase
phenomena does not depend in any simple way on that expected from
either anisotropy term alone.

\section{\label{sec:level1} High Frequency EPR Experiments}

High frequency (40 to 200 GHz) single crystal Electron
Paramagnetic Resonance (EPR) measurements were carried out using a
millimeter-wave vector network analyzer (MVNA) and a high
sensitivity cavity perturbation technique; this instrumentation is
described elsewhere \cite{Mola}. Temperature control in the range
from 2 K to room temperature was achieved using a variable-flow
cryostat. The magnetic field was provided by a horizontal
superconducting split-pair magnet with vertical access, enabling
angle dependent studies ($<$0.1$^\circ$ resolution) and
approximate alignment of the single crystal \cite{Hill3}. In order
to make accurate comparisons with the magnetization studies
presented in the preceding sections, all data presented in this
section were performed on a single deuterated crystal of
Mn$_{12}$-ac (d-Mn$_{12}$-ac), having approximate dimensions
$1\times0.3\times0.3$ mm$^3$. This crystal was selected from a
batch of samples which had previously been removed from the mother
liquor and stored for 1 year in a refrigerator (at 5$^\circ$
Celsius) prior to the measurement. This particular batch was grown
using standard methods \cite{Hendrickson}, albeit in a completely
independent synthesis from the samples used for the magnetization
measurements. The sample was separately cooled under vacuum from
room temperature at 5 K/minute in one of two orientations for
field rotation in i) the $x-y$ plane, and ii) a plane
perpendicular to the $x-y$ plane. In the former case, the sample
was mounted on the side wall of a cylindrical TE011 cavity (center
frequency = 53.1 GHz, Q$\sim$20,000) with its easy axis parallel
to the cavity axis such that the microwave field $H_1$ was aligned
parallel to the sample's easy axis (and, therefore, perpendicular
to the applied DC field). In the latter case, the sample was
mounted on the end plate of the same cavity, and DC field rotation
was carried out for angles close to the $x-y$ plane (within
15$^\circ$), with the microwave $H_1$ field again parallel to the
sample's easy $z$-axis. Field sweeps were restricted to 6.6 T due
to limitations of the split-pair magnet. As will become apparent,
the data obtained for the d-Mn$_{12}$-ac are in qualitative
agreement with earlier published results obtained for the
hydrogenated Mn$_{12}$-ac \cite{Hill2,Hill3}.

\subsection{Magnetic symmetry measurements in the high-field limit}

In the preceding sections, it has been shown how the Landau-Zener
method may be applied to SMMs in order to determine very weak
transverse terms in Eq. (1). Moreover, the Landau-Zener method
allows one to select molecules, based on the tunneling rates of
the different species. Subsequently, by performing angle dependent
studies on each sub-species, one can deduce the underlying
symmetries of the dominant tunneling matrix elements. While this
method is extremely powerful, it is evident from the discussion in
section II.C that, for systems with multiple sources of transverse
anisotropy (intrinsic and extrinsic), deconvolution of the
different contributions to the tunnel splittings can be
problematic, i.e. in Figs. 6 to 9 it is seen that competing $E$
and $C$ terms results in a competition between the two-fold and
four-fold symmetries of these interactions. The reason for this
competition is that, at low-fields, the $E$ and $C$ terms operate
in different high orders of perturbation theory, thereby resulting
in a complicated interaction between the two perturbations.

At first sight, it is not obvious how EPR experiments, conducted
in the GHz frequency range, could shed new light on the nature of
transverse interactions which manifest themselves as miniscule
tunnel splittings of order 10$^4$ Hz at low-fields. However, the
so-called "tunnel-splittings" are measured by the Landau-Zener
method at low fields wherein the transverse terms operate in very
high orders of perturbation theory within an $m_z$ basis (see Eq.
(3)), where the quantization axis is defined by the global
easy-axis of the crystal. For high magnetic fields applied in the
transverse direction, such a picture is no longer valid due to the
conflicting symmetries imposed by the crystal field and the
applied field. Herein lies the beauty of the high-field EPR
technique. By applying a sufficiently strong transverse field, one
can reach a limit in which the appropriate basis of spin states is
defined by a quantization axis parallel to the applied field, i.e.
the $x$-direction. In such a limit, transverse zero-field
interactions operate in zeroth-order. Consequently, their effects
may be rather strong. As an illustration of this point, consider
the simplest zero-field Hamiltonian:
\begin{equation}
\label{eq.15}{\cal {H}}=-DS_z^2+E(S_x^2-S_y^2)-g\mu _BH_xS_x\;,
\end{equation}
Making a substitution for $S_z^2$ in terms of $(S_x^2+S_y^2)$, one
obtains:
\begin{equation}
\label{eq.16}{\cal
{H}}={\frac{1}{2}}(D+3E)S_x^2+{\frac{1}{2}}(D-E)(S_y^2-S_z^2)-g\mu
_BH_xS_x\;,
\end{equation}
which can be re-written as:
\begin{equation}
\label{eq.17}{\cal {H}}={\frac{1}{2}}(D+3E)S_x^2-g\mu
_BH_xS_x+{\cal {H'_T}}\;,
\end{equation}
This equation is diagonal in $S_x$, and has the same form as Eq.
(15). Therefore, to lowest order, the high-field eigenvalues will
be given by:
\begin{equation}
\label{eq.18}\epsilon(m_x)\approx{\frac{1}{2}}(D+3E)m_x^2-g\mu
_BH_xm_x\;,
\end{equation}
Similar arguments hold for higher order transverse terms. Thus,
the transverse high-field EPR spectra provide perhaps the most
direct means of measuring these transverse terms. Indeed, this
represents one of the more illustrative examples of the importance
of high-field EPR as a spectroscopic tool for studying quantum
magnetism. Not only does the effect of the zero-field transverse
terms shift to zeroth order, but the symmetry of such interactions
is also preserved. This is best illustrated using the same example
as above, with the field applied along the $y$-axis instead of the
$x$-axis. In this case, Eq. (15) may be re-written:
\begin{equation}
\label{eq.19}{\cal {H}}={\frac{1}{2}}(D-3E)S_y^2-g\mu
_BH_yS_y+{\cal {H''_T}}\;,
\end{equation}
giving
\begin{equation}
\label{eq.20}\epsilon(m_y)\approx{\frac{1}{2}}(D-3E)m_y^2-g\mu
_BH_ym_y\;,
\end{equation}
Thus, from Eqs. (18) and (20), it is apparent that the influence
of the (E) term changes sign upon rotating the applied field from
the $x$-axis of the $E$-tensor to the $y$-axis. This two-fold
behavior is not unexpected; indeed, it is obtained also from the
exact diagonalization calculations shown in Figs. 6 to 8, which
correspond to the high-field/frequency limit discussed here.
However, unlike lower-field calculations, the 2$^{nd}$ and
4$^{th}$ order interactions decouple completely at high-fields, as
do other transverse interactions. Thus, one may consider their
effects completely independently. This point is illustrated in
Fig. 8, where it can be seen that the effect of an intrinsic
fourth-order anisotropy is to produce a
ground-to-first-excited-state splitting ($\Delta_{k=0}$) which
oscillates as a function of the field orientation within the
hard/medium plane, with a periodicity of 90$^\circ$. The
disorder-induced rhombic anisotropy, meanwhile, has no affect on
this four-fold behavior. It simply causes a two-fold modulation
of the tunnel splitting, which superimposes onto the four-fold behavior caused
by the fourth-order interaction. Because of the complete
independence of these effects, one can in principle determine any
misalignment, $\beta$, between the hard axes associated with the
second and fourth order transverse anisotropies, as illustrated in
Fig. 8.

In order to make direct comparisons with other spectroscopic
studies (e.g. neutron \cite{Mirebeau} and EPR
\cite{Barra,Hill,Cornia}) we re-write the Hamiltonian of Eq. (1)
in the following form:
\begin{equation}
\label{eq.21}{\cal
{\hat{H}}}=D^\prime[\hat{S}_z^2-{\frac{1}{2}}\hat{S}(\hat{S}+1)]+E\hat{O}_2^{2'}+B_4^0\hat{O}_4^0+B_4^4\hat{O}_4^4+{\cal
{\hat{H}}_Z}\;
\end{equation}
where ${\cal {\hat{H}}_Z}=\mu
_B\hat{B}\cdot\hat{\hat{g}}\cdot\hat{S}$ is the Zeeman term due to
the applied magnetic field, $E\hat{O}_2^{2'}$ is the first term in
eq. (5), and $\hat{O}_B^A$ are the Steven's operators, of order
$B$ in the spin operators and possess $A$-fold symmetry (i.e.
$\hat{O}_4^4\equiv{\frac{1}{2}}(\hat{S}_+^4+\hat{S}_+^4)$). The
uniaxial parameter $D^\prime$ is not exactly the same as the $D$
parameter in Eq. (1) [or Eqs. (15)-(20)]. This is due to the
occurrence of an $\hat{S}_z^2$ term in the $\hat{O}_4^4$ Steven's
operator, i.e. the presence of a significant fourth-order axial
anisotropy has the effect of renormalizing the quadratic
$m$-dependence of the barrier \cite{Dprime,Pederson3}.\\
\begin{figure}
\begin{center}\includegraphics[width=8.6cm]{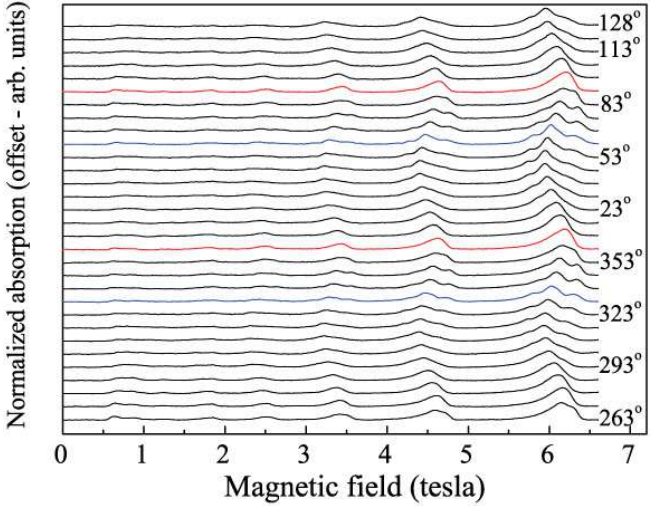}
\vspace{-2 mm}\caption{Microwave absorption obtained for different
field orientations within the hard ($xy$-) plane of the sample;
the temperature was 15 K and the frequency was 51.3 GHz in every
case. The peaks in absorption correspond to EPR. The data were
obtained at $7.5^\circ$ intervals, where the angle $\phi$ refers
to the field orientation relative to one of the flat edges of the
square cross section of the sample. The resonances have been
labeled according to the scheme described in ref. \cite{Hill3}.
The red traces correspond to field orientations approximately
parallel to the hard/medium axes of the $E$ tensor, i.e.
orientations corresponding to the maximum splitting of the low and
high field shoulders. The blue traces correspond to orientations
of the hard axes of the $B_4^0$ tensor.}\vspace{-6 mm}
\end{center}
\end{figure}

Fig. 29 displays the raw microwave absorption obtained for
different field orientations within the hard ($x-y$) plane of the
sample; the temperature was 15 K and the frequency was 51.3 GHz in
every case. The peaks in absorption correspond to EPR. The data
were obtained at 7.5$^\circ$ intervals, where the angle $\phi$
refers to the field orientation relative to one of the flat edges
of the square cross section of the sample. The resonances have
been labeled according to the scheme described in ref.
\cite{Hill3}. For fields applied approximately parallel to the
hard plane, only $\alpha-$resonances are observed
($\beta-$resonances appear for field rotation away from the hard
plane, see below). The highest field peak, $\alpha8$, corresponds
to an excitation between levels which evolve from the zero field
$m_z=\pm9$ zero-field doublet. The transition from the ground
state, $m_z=\pm10$ ($k=0$ resonance), is not observed within the
available field range for these experiments; at 51 GHz, its
expected position is at $H_T\sim$9 T. For a detailed discussion of
the resonance labeling scheme, as well as the temperature,
frequency, field and field orientation dependence of the EPR
spectra for Mn$_{12}$-ac, refer to \cite{Hill3}.
\begin{figure}
\begin{center}\includegraphics[width=8.6cm]{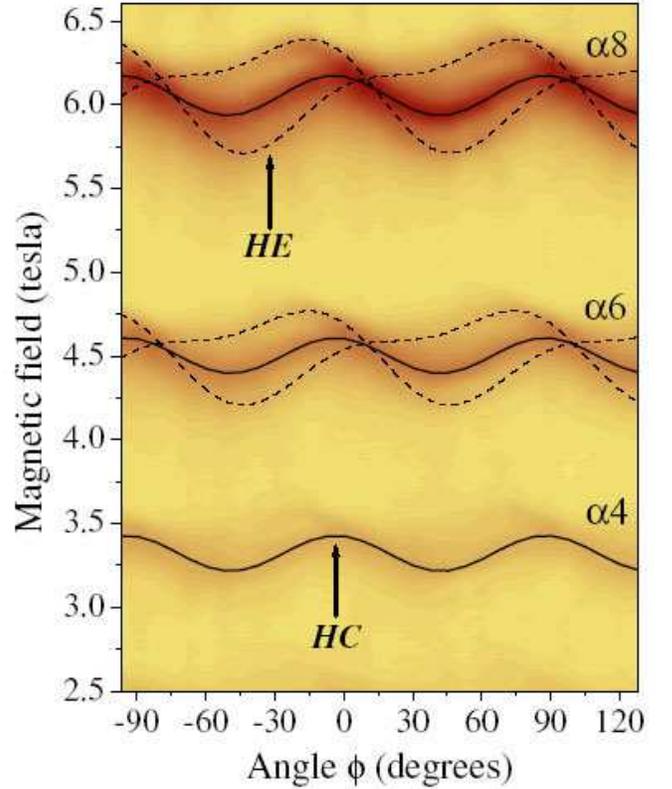}
\vspace{-2 mm}\caption{Color contour plot of the absorption
intensity (see Fig. 29) versus magnetic field strength and the
azimuthal angle $\phi$; the darker shades correspond to stronger
absorption. Superimposed on the absorption maxima are fits to the
$\phi$-dependence of the central positions of each peak (solid
blue lines), as well as fits to the positions of the shoulders
(solid red lines). The horizontal bars are fits to Eq.~(21) (see
main text for explanation). The approximate orientations of the
hard axes corresponding to the $E$ (HE) and $B_4^4$ (HC) tensors
are indicated. The open circles represent recent data points
obtained for h-Mn$_{12}$-ac~\cite{Hill4}.}\vspace{-6 mm}
\end{center}
\end{figure}

Immediately apparent from fig. 29 is a four-fold variation in
the positions of each cluster of resonances ($\alpha$8, $\alpha$6,
etc.). Note that each of the resonances exhibit fine structures,
which also depend on the field orientation $\phi$; these are
related to the disorder in the crystal, which we discuss
further below. The four-fold shifts are due to the intrinsic
fourth-order transverse anisotropy $B_4^4\hat{O}_4^4$, as has
previously been established for h-Mn$_{12}$-ac \cite{Hill2}. Fig.
30 shows a color contour plot of the absorption intensity versus
magnetic field strength and the azimuthal angle $\phi$.
Superimposed on the absorption maxima are two kinds of fit to the
$\phi$-dependence of each peak. The solid blue curves were
obtained simply by fitting
the positions of the central peaks in Fig. 29 with pure sine functions having
four-fold periodicity. The fits represented by horizontal bars, meanwhile, were
obtained by exact diagonalization of the Hamiltonian matrix, assuming accepted
values for the zero-field parameters $D^\prime$ and
$B_4^0$ ($D^\prime$ = $-0.455$ cm$^{-1}$ and $B_4^0$ = 2$\times 10^{-5}$~ cm$^{-
1}$).
The fourth-order transverse parameter
$B_4^4=3.2\times10^{-5}$ cm$^{-1}$ is the only unknown
parameter in the fit. $D^\prime$ and $B_4^0$ were verified independently
from easy axis measurements, and all peak positions are
consistent with a single value of $B_4^4$. We note that this value
is in precise agreement with that found for h-Mn$_{12}$-ac, as is
to be expected for this fourth-order interaction which is related
to the intrinsic symmetry of the Mn$_{12}$O$_{12}$ molecule. From
the maxima and minima in the peak position shifts induced by the
$B_4^4\hat{O}_4^0$ term, we estimate that the hard and medium
directions of this intrinsic crystal field interaction are
oriented at $\phi_{HC}=-4.5^\circ\pm5^\circ$ ($+i90^\circ$) and
$\phi_{MC}=41.5^\circ\pm5^\circ$
($+i90^\circ$) relative to the square edges of a typical single crystal
sample.\\

Next we turn to the angle-dependent fine structures which are very
apparent in the ranges $\phi=300^\circ-330^\circ$ and
$\phi=30^\circ-60^\circ$ in fig. 29. The first point to note is
the fact that we see shoulders on both the high and low-field
sides of the main peaks in these angle ranges. This contrasts the
situation found from our earlier studies of h-Mn$_{12}$-ac~\cite{Hill2}, where
only high-field
shoulders were observed. We comment on these differences at the end of this
section.
We first discuss the origin of the angle dependence of the shoulders, which are
very apparent in the ranges $\phi = 300^\circ - 330^\circ$ and $\phi = 30^\circ
- 60^\circ$
in fig. 29; a more in-depth discussion can be found in refs. \cite{Hill2,Hill3}.
The high- and low-field shoulders are due to the $n = 1$ and $n =
3$ Mn$_{12}$-ac hydrogen-bonding variants in Cornia's solvent
disorder model \cite{Cornia}.
These variants, which comprise roughly $50\%$ of the total molecules, are
thought to possess a significant rhombic anisotropy due to the reduced symmetry
of the surrounding hydrogen bonded acetic acid solvent molecules.
The second-order operator, $\hat{O}_2^2$, associated with the
rhombic distortion gives rise
to two-fold behavior, as was clearly demonstrated from the magnetic measurements
in the previous section of this paper. It is important to recognize that the
disorder is discrete, since the acetic acid can only bond at four positions on
the Mn$_{12}$ molecule. Thus, one expects only two hard directions associated
with the $n = 1$ and $n = 3$ low symmetry variants, which are separated by
$90^\circ$. In HFEPR, the $\hat{O}_2^2$ operator causes shifts in the peak
positions. When the applied transverse field is parallel to one of the hard
axes, it is obviously perpendicular to the other, which causes shifts in the
HFEPR intensity to both the low- and high-field sides of the central peaks,
hence the shoulders. When the field is applied in between these two directions,
the EPR intensity due to the low-symmetry variants collapses into the central
portion of the peak, hence the disappearance of the shoulders every $90^\circ$
(see figs. 29 and 30). In actual fact, this apparent four-fold behavior reflects
the two-fold nature of the rhombic distortion caused by the hydrogen bonding
acetic acid molecules, with the EPR intensity for a given variant shifting from
the low (high) to the high (low) field side of the main peak every $90^\circ$,
i.e. the periodicity is actually $180^\circ$. This two-fold behavior is then
superimposed on the intrinsic four-fold periodicity, as can be seen from the red
curves in fig. 30.

An important point to note from the hard-plane rotation data is
the phase shift between the four-fold modulation of the central
peak position (blue curves in fig. 30) and the two-fold shifts of
the low and high field shoulders (red curves in fig. 30). Indeed,
this is one of the main points of this article. As discussed in
section II.C, this difference indicates a $\sim27^\circ\pm3^\circ$
mis-alignment of the $\hat{O}_2^2$ and $\hat{O}_4^4$ tensors, as
was originally suspected for h-Mn$_{12}$-ac
\cite{delBarco2,Hill2}. The present study provides further support
for this finding, thereby illustrating the remarkable differences
between the global and local symmetries of Mn$_{12}$-ac.
\begin{figure}
\begin{center}\includegraphics[width=8.6cm]{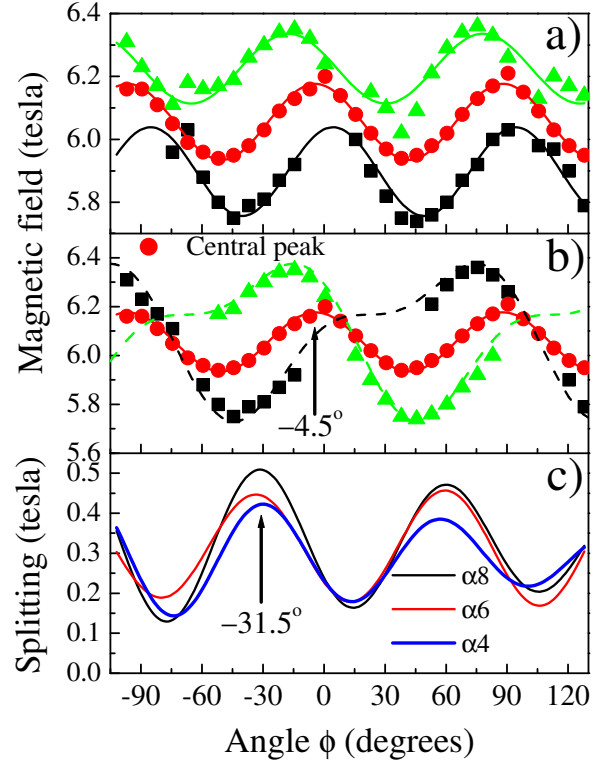}
\vspace{-2 mm}\caption{a) Hard-plane angle ($\phi$) dependence of
the splitting of the high and low-field shoulders, for the
$\alpha$8 peak. This figure may be compared with Fig. 2 of ref.
\cite{Hill}, which displays similar data for the h-Mn$_{12}$-ac
complex. In (a), the positions of the low- (black) and high-field
(green) shoulders on $\alpha$8 have been separately fit with sine
functions. The difference between these fits is displayed in (c),
together with similar curves generated by the same procedure for
peaks $\alpha$6 and $\alpha$4. The splittings plotted in (c) are a
measure of the shifts caused by the disorder-induced rhombic term
for the low-symmetry Cornia variants \cite{Cornia} (see also Fig.
8); all curves are in agreement ($\pm2^\circ$) as to the
orientation of the hard/medium two-fold directions. In (b), the
data in (a) are plotted in such a fashion as to illustrate the
real two-fold nature of the angle dependence of the high- and
low-field shoulders. The approximate orientation
($\phi=-4.5^\circ$) of one of the hard four-fold axes (HC) is
indicated in (b), and the approximate orientation
($\phi=-31.5^\circ$) of one of the hard/medium two-fold axes (HE)
is indicated in (c).}\vspace{-6 mm}
\end{center}
\end{figure}

In order to make quantitative comparisons between the
disorder-induced effects in h-Mn$_{12}$-ac and d-Mn$_{12}$-ac, we
performed a single fit to each of the peaks in fig. 30 via exact
diagonalization of the Hamiltonian matrix [Eq. (21)]. This fit is
represented by the horizontal bars in fig. 30. Our procedure
obviously takes into account the mis- alignments of the
$\hat{O}_2^2$ and $\hat{O}_4^4$ tensors, as described in section
II-C. Thus, the employed Hamiltonian is subtly different from the
standard form used by most spectroscopists, which may explain
slight differences in the obtained Hamiltonian parameters. The
only free parameters in the fit were then the $E$ and $B_4^4$
coefficients (corresponding to the $\hat{O}_2^2$ and $\hat{O}_4^4$
tensors), for which we obtained the values $\pm0.014(2)$~cm$^{-1}$
and $\pm  3.2(5) \times 10^{-5}$~cm$^{-1}$ respectively. The $E$
value is significantly larger than the one obtained in earlier
experiments ref.~\cite{Hill2}). We attribute some of this
difference to the modified Hamiltonian used in our more recent
fits, which takes into account the mis-alignments of the
$\hat{O}_2^2$ and $\hat{O}_4^4$ tensors. However, much of the
difference appears to be real. The larger $E$-value found from the
present study is somewhat surprising. However, as will be seen
below, more recent measurements on a very fresh h-Mn$_{12}$-ac
sample are in excellent agreement with the value of
$\pm0.014(2)$~cm$^{-1}$ found from this study (open blue circles
in fig. 30). Thus, the difference is likely related to sample
quality and/or solvent loss. A tell-tale sign of the higher sample
quality is the observation of both high- and low-field shoulders
on the main EPR peaks. In contrast, only high-field shoulders were
seen in the earlier experiments on h-
Mn$_{12}$-ac~\cite{Hill2,Hill3}. In fact, the absence of a
low-field shoulder is discussed at some length in ref.
\cite{Hill3}, where it is shown that this peak is unresolved from
the broad low-field tail associated with the central portion of
the peak. The reason for the asymmetry between the high and
low-field shoulder is related to easy axis tilting caused by the
solvent disorder (discussed in the next sub-section). Although
both shoulders are seen in the present study, a clear asymmetry
can be seen from the data in fig. 29. We suspect that the samples
used in the earlier experiments may have suffered significant
solvent loss, either upon cooling from room temperature under high
vacuum, or simply as a result of being stored in air for more than
1 year prior to the measurements. Indeed, variations in the
$D$-strain measured in different Mn$_{12}$-ac samples has
previously been reported by us \cite{Hill1}. Increased $D$-strain
leads to broader EPR lines, thus probably explaining why the
shoulders are clearly resolved in the present investigation, but
not in refs.~\cite{Hill2,Hill3}. Without two well resolved
shoulders, it is likely that the $E$-strain was under-estimated in
ref.~\cite{Hill2}, or it could simply be that the $E$-strain is
weaker in samples that suffer significant solvent loss. Based on
more recent experience working with SMMs containing considerably
more volatile solvents, we have recently developed sample handling
procedures which minimize solvent loss, e.g. encapsulating samples
in oil prior to cooling under atmospheric helium gas. The
differences between the present measurements and those reported in
refs.~\cite{Hill2,Hill3} highlight the importance of sample
handling. Indeed, it is likely that Mn$_{12}$-ac samples prepared
by different groups, and studied by different techniques, exhibit
significant differences in their solvent content, resulting in
subtly different conclusions concerning the quantum dynamics.
Solvent loss probably also provides an explanation for the widths
of the distributions of tunnel splittings found in the magnetic
relaxation experiments described in section III. For this reason,
it will be advantageous to prepare Mn$_{12}$-ac SMMs which do not
exhibit such a dramatic dependence on solvent content.

\subsection{Easy-axis tilting}

In this section, we briefly present the results of measurements
for field rotations away from the hard plane in order to
illustrate the presence of a small distribution of tilts of the
easy axes of the d-Mn$_{12}$-ac molecules. In the following
figures, the polar angle $\theta$ represents the angle between the
applied DC magnetic field and the global easy axis of the crystal,
i.e. $\theta=90^\circ$ indicates the hard/medium-plane direction.
Field rotation was performed in a plane approximately parallel to
one of the large flat surfaces of the needle-shaped crystal (i.e.
$\phi=0$, or 90$^\circ$, etc.). A more extensive discussion of the
analysis and interpretation of the results of similar experiments
for the h-Mn$_{12}$-ac complex have been published elsewhere
\cite{Hill3}. Consequently, we present only data for
d-Mn$_{12}$-ac in this article, and discuss the implications
without including detailed simulations, which will be published
elsewhere \cite{Hill4}.
\begin{figure}
\begin{center}\includegraphics[width=8.6cm]{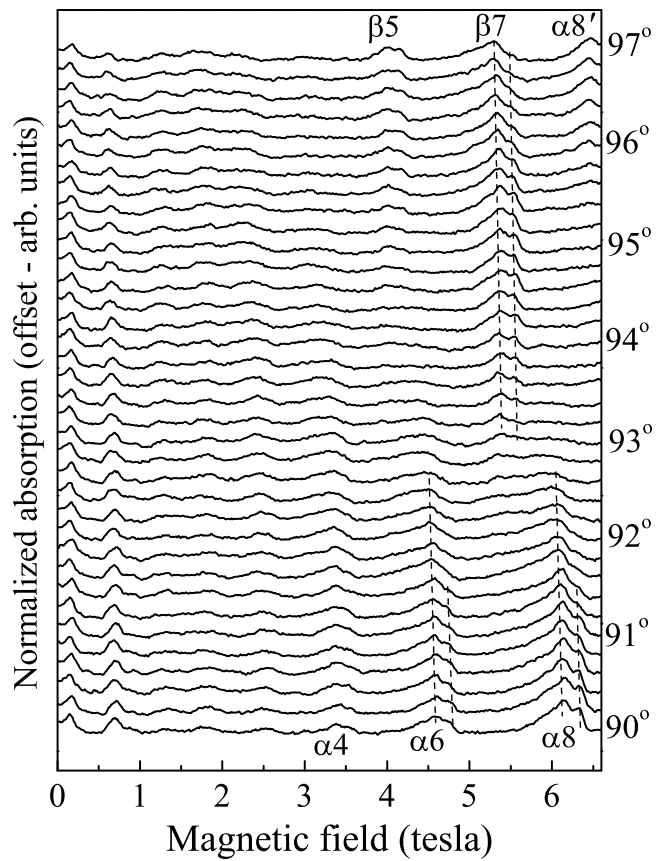}
\vspace{-2 mm}\caption{Microwave absorption spectra obtained at
$0.2^\circ$ intervals over the range from $\theta=90^\circ$ to
$97^\circ$ ($\phi\sim5^\circ$); the temperature was 15 K and the
frequency was 51.3 GHz in every case, and the traces are offset
for clarity. The peaks in absorption correspond to EPR, and the
labeling is discussed in ref. \cite{Hill3}. See main text for
discussion of the data.}\vspace{-6 mm}
\end{center}
\end{figure}

Fig. 32 shows a series of absorption spectra obtained at
$0.2^\circ$ intervals over the range from $\theta=90^\circ$ to
$97^\circ$. Again, the peaks in absorption correspond to EPR, and
the labeling is discussed in ref. \cite{Hill3}. The quality of the
data is noticeably poorer than the data in fig. 29, and is due to
the positioning of the sample at a different location in the
cavity, where the geometry of the electromagnetic fields are not
quite optimal for EPR. Fig. 33 displays (a) a gray scale contour
map and (b) a 3D plot representing the same data shown in fig. 32
(the darker colored regions correspond to stronger EPR
absorption), albeit for a wider range of angles (up to
$\theta=105^\circ$). As discussed at great length in ref.
\cite{Hill3}, the transverse-field EPR spectra for Mn$_{12}$-ac
exhibit unusual selection rules for frequencies below 90 GHz: two
series of resonances are separately observed (labeled $\alpha$ and
$\beta$) as the magnetic field is tilted away from the hard plane.
These selection rules are extremely sensitive to the field
orientation for angles close to the hard plane. Thus, deviations
from the behavior predicted by the giant spin model (Eq. (1))
provide evidence for tilts of the molecules.
\begin{figure}
\begin{center}\includegraphics[width=8.6cm]{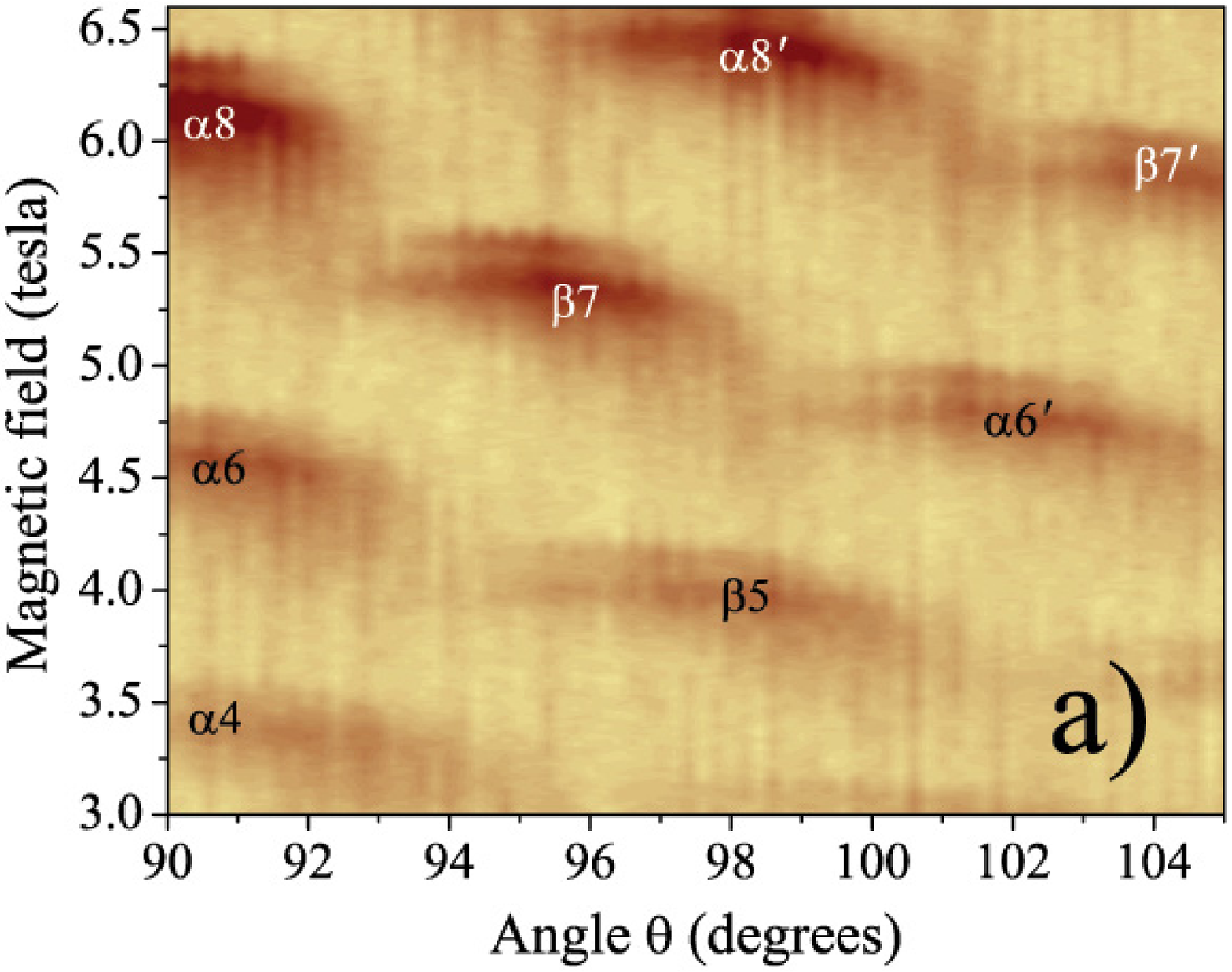}
\end{center}
\end{figure}
\begin{figure}
\begin{center}\vspace{-8 mm}\includegraphics[width=8.6cm]{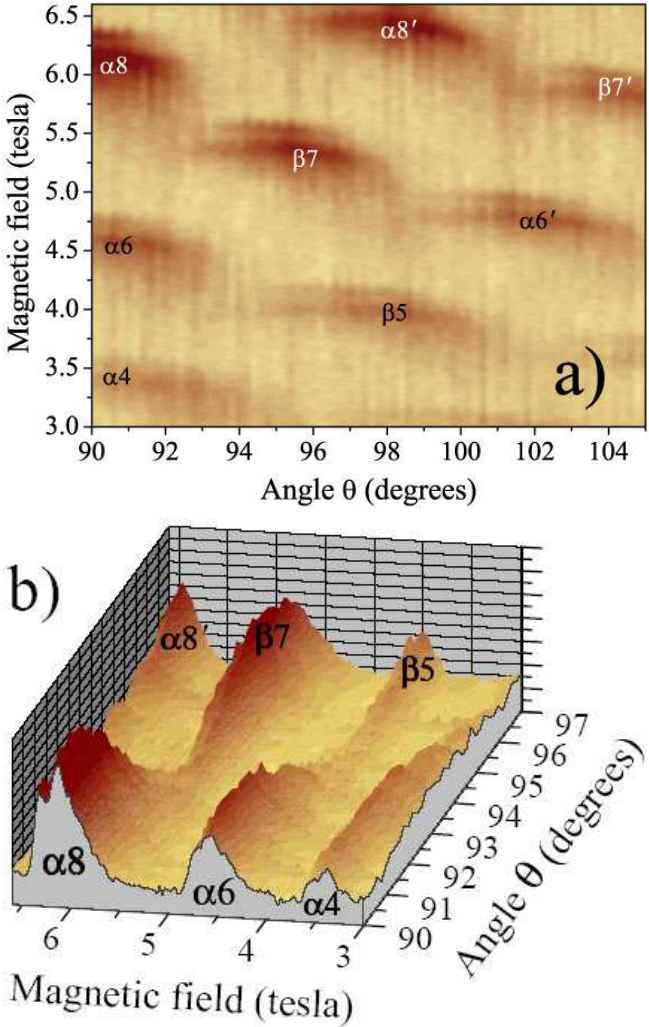}
\vspace{-2 mm}\caption{(a) A color contour map and (b) a 3D color
surface plot representing the same data shown in Fig. 32 (the
darker colored regions correspond to stronger EPR absorption). We
plot the data in this manner for direct comparison with
simulations shown in Fig. 6. The significant overlap of the
$\alpha$ and $\beta$ resonances suggests tilting of the molecules.
See main text and ref. \cite{Hill3} for discussion.}\vspace{-6 mm}
\end{center}
\end{figure}

51.3 GHz simulations of the EPR spectra for an idealized
Mn$_{12}$-ac sample are displayed in fig. 34 $-$ both a color
contour map for the full $\theta=90^\circ$ to $105^\circ$ range
(fig. 34.a), and a 3D view for the $\theta=90^\circ$ to $97^\circ$
range (fig. 34.b). These 15 K simulations were generated using
the accepted Hamiltonian parameters given above. A Gaussian line shape
was employed with a line width typical for a single
solvent-disorder variant in Cornia's model (see ref.
\cite{Hill3}). While this line shape/width is clearly quite
different from the experimental data, which is complicated by the
contributions of several Mn$_{12}$-ac variants, we note that the
choice of line width/shape does not in any way affect the
following analysis. The simulations indicate a range of about
$1.6^\circ$ (from $91.8^\circ$ to $93.4^\circ$) over which the
EPR intensity associated with both the $\alpha$8 and $\alpha$7 transitions
is negligible. Thus, for a perfect crystal, without any disorder,
one should expect a similar behavior in the actual EPR spectra.
However, careful inspection of fig. 32 and 33 indicates a
significant overlap of the $\alpha$8 and $\beta$7 peaks in the
$92^\circ$ to $93^\circ$ range. These two facts point to a spread
in the orientations of the magnetic axes of the molecules, with a
cut-off of at $\sim 1.3^\circ$ away from the global directions,
i.e. we predict that, on average, the magnetic easy axes of the
low-symmetry (disordered) molecules are tilted away from the
crystallographic $z$-axis, and that the distribution extends roughly
$1.3^\circ$. Once again, a very
similar behavior has been observed for h-Mn$_{12}$-ac in ref.
\cite{Hill3}, albeit that the distribution extends to $\sim 1.7^\circ$. We note
that the h-Mn$_{12}$-ac experiments were
conducted using a rotating cavity \cite{Takahashi} to higher
magnetic fields, where the experimental deviation from the
simulation is even more pronounced. Thus, easy axis tilting
appears to be a general feature in Mn$_{12}$-ac and, as discussed
earlier in this article, this likely provides an explanation for
the lack of selection rules in the magnetization steps observed
from hysteresis experiments. The slightly broader tilt
distribution for h-Mn$_{12}$-ac may be related to the increased
solvent loss.

Finally, we comment on the correlation between the
disorder-induced rhombic anisotropy and the easy axis tilting. It
has been shown by Park et al. \cite{Pederson}, that the magnetic
anisotropy tensor absolutely determines the orientations of the
principal magnetic axes. In other words, one expects the effects
of the solvent disorder to be accompanied by local easy-axis
tilting. This was also pointed out by Cornia et al. \cite{Cornia}
in their X-ray analysis of Mn$_{12}$-ac. We have subsequently
shown this to be the case from EPR studies of h-Mn$_{12}$-ac
\cite{Hill3}, where it was shown that the high-field fine
structures in the transverse EPR spectra displayed a different
angle dependence compared to the main peaks. These findings
suggested that each solvent disorder variant has a distinct angle
dependence and, therefore, a distinct tilting behavior, i.e. the
tilting and the anisotropy are correlated. This is again apparent
for d-Mn$_{12}$-ac from fig. 33.a, where the shoulders are visible
as narrow horizontal streaks on the high field sides of each of
the main peaks. It is quite evident that the narrow streaks (i.e.
the shoulders) span a narrower angle range compared to the main
peaks. In fact, the high field shoulders on the $\alpha_8$ and
$\alpha_7$ peaks exhibit a considerable range where neither is
observed. Indeed, this range corresponds almost exactly to the
$1.6^\circ$ found from the simulations (fig. 34). It is important
to note that this local tilting of the magnetic axes (caused by
local symmetry lowering), is quite different from physical tilting
of the molecules caused e.g. by strains in the sample.
\begin{figure}
\begin{center}\includegraphics[width=8.6cm]{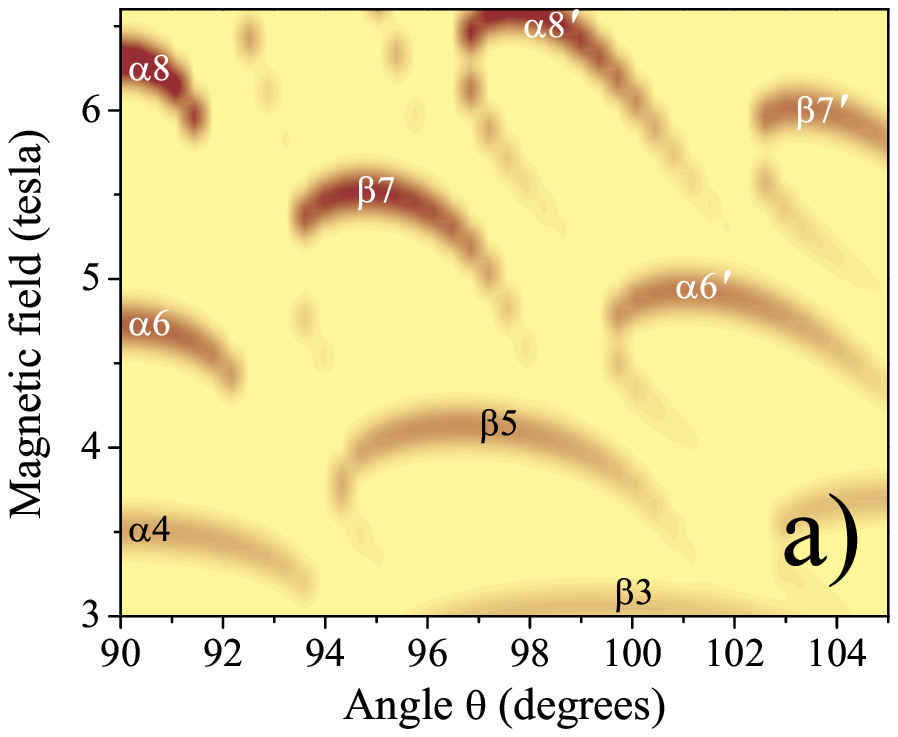}
\end{center}
\end{figure}
\begin{figure}
\begin{center}\vspace{-8 mm}\includegraphics[width=7.5cm]{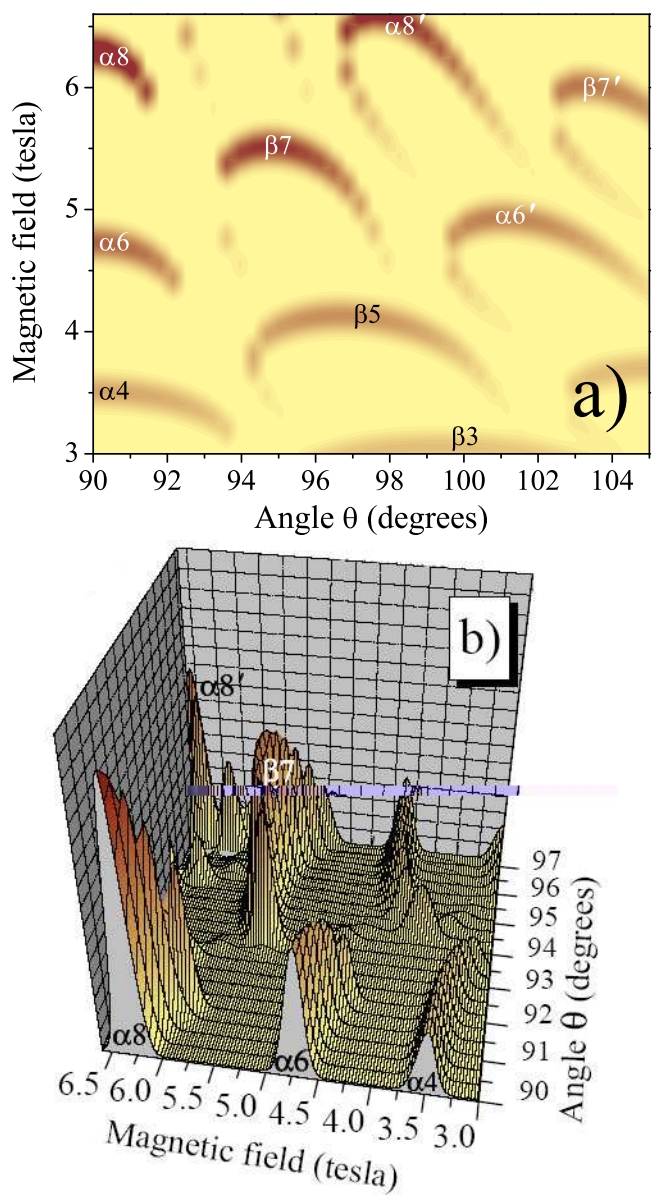}
\vspace{-2 mm}\caption{(a) A color contour map and (b) a 3D color
surface plot representing simulations of the data in Fig. 33,
assuming that all molecules are aligned, i.e. no tilting. The
simulations were generated using the Hamiltonian parameters given
in the text, and the temperature and frequency are 15 K and 51.3
GHz respectively.}\vspace{-6 mm}
\end{center}
\end{figure}

Comparisons between Figs. 32, 33 and 34 indicate that the angle
dependence of the high-field shoulders agrees very well with the
simulations. The reason for these differences between the main
peaks and the shoulders is explained in ref. \cite{Hill3} as being
due to discrete easy axis tilting, wherein the tilting is confined
to two orthogonal planes defined by the hard and medium directions
of the associated disorder-induced $\hat{O}_2^2$($E$) zero-field
tensor. Essentially, the high-field shoulders on the EPR peaks are
due to molecules which are tilted in a plane which is
approximately perpendicular to the plane of rotation of the
applied magnetic field. Consequently, the tilts due to these
molecules do not project onto the field rotation plane, i.e. the
experiment is insensitive to tilts in this direction. Meanwhile,
the low-field tail of the EPR peaks is due to molecules which tilt
in the orthogonal plane. For these molecules, the tilts have a
maximum projection on to the field rotation plane, i.e. the
experiment is maximally sensitive to tilts in this direction. The
present studies support the findings of the original study for
h-Mn$_{12}$-ac \cite{Hill3}, providing further confirmation for
the discrete tilting idea.\\

\section{\label{sec:level1} Conclusions}

In summary, these experiments provide a comprehensive
understanding of the factors that influence the symmetry of MQT in
Mn$_{12}$-ac, the first and most widely studied SMM.
Interestingly, the reason so much attention has focused on
Mn$_{12}$ has been the high global symmetry of single crystals, as
most known SMMs have lower site symmetry. The data presented here
show that disorder lowers the symmetry locally and leads to an
intricate interaction between transverse anisotropy terms: the
first associated with disorder and the second intrinsic to an
``ideal'' Mn$_{12}$-molecule. While at first sight a nuisance, these
complexities have been interesting from many perspectives. From
experiment, they have been a challenge to understand and
characterize, and have required combining unique advanced and
sensitive magnetic techniques that have been developed by the
authors over many years. In particular, this research involved
combining single crystal high-frequency and high-field EPR and
low-temperature magnetometry, both with arbitrarily directed
applied magnetic fields. The results obtained by magnetic
measurements of MQT have had implications for EPR studies and
vice-versa. From a theoretical perspective, given a structural
model of the molecule and solvent environment, density functional
theory has been able to capture many of the features observed in
these experiments, including the magnitudes and form of the
transverse magnetic anisotropies and the easy axis tilts
\cite{Pederson}. This includes the angle between the 2nd order and
4th order transverse anisotropies, which has been central to
understanding the combined data set, as well as the easy axis
tilts. This research thus represents an important milestone in
our understanding of the factors that influence MQT in SMMs.\\

Acknowledgments. This research was supported by NSF (Grant Nos.
DMR-0103290, 0114142, 0239481 and 0315609). S. H. acknowledges
Research Corporation for financial support.


\begin{references}
\bibitem{Hendrickson} d-Mn$_{12}$-acetate was synthesized by following the
standard procedure for preparing h-Mn$_{12}$-acetate, but using
D$_2$O and deuterated acetic, CD$_3$COOD.
\bibitem{note1} In the result presented in ref. \cite{delBarco2} the one-fold
contribution of this measurement was displaced by 180 degrees.
This corresponds to a mistake that was due to the fact that for
this experiment we took $\phi=-180^\circ$ (that corresponds to
negative orientation of one of the coils of the superconducting
vector magnet) as the origin of rotation while for all the others
the origin was $\phi=0$ (positive orientation of the same coil)
and we did not correct it in the data.
\bibitem{note2} We studied the behavior of the MQT probability with a transverse
field applied along the directions of a maximum and a minimum of
the twofold MQT probability response because our initial
interpretation of the results in ref. \cite{delBarco2} neglected
the effect of incommensurate transverse anisotropies and we
assumed that those were the directions of the hard and the medium
anisotropy axes associated with $E(S_x^2-S_y^2)$. As we explain in
this article, these directions, in fact, correspond to medium
anisotropy axes of the fourth anisotropy term and none of them are
along the hard anisotropy axes of the second anisotropy term.
\bibitem{Dprime} The $\hat{O}_4^0$ Stevens operator has the form
$35\hat{S}_z^4+[25-30S(S+1)]\hat{S}_z^2$. Thus, a finite $B_4^0$
term results in an additional quadratic contribution to the
anisotropy barrier. Consequently, the axial crystal field
parameters $D$ (Eq.~1) and $D^\prime$ (Eq.~21) are inequivalent in
situations where $B$ (and, therefore, $B_4^0$) are finite; $B$ and
$B_4^0$ are inequivalent also. $D$ and $D^\prime$, as well as $B$
and $B_4^0$, also have opposite signs. The appropriate
transformation between the two parameter sets are as follows:
$D=(3275\times B_4^0)-D^\prime$ and $B=-35\times B_4^0$.
\end{references}
\end{document}